\documentclass[aps,prd,preprint,superscriptaddress,preprintnumbers,nofootinbib]{revtex4-1}
\pdfoutput=1

\usepackage{graphics}

\usepackage[dvips]{graphicx}
\usepackage{mathrsfs}
\usepackage{amssymb}
\usepackage{amsmath}
\usepackage{verbatim}
\usepackage{float}
\usepackage{slashed}
\usepackage{soul}
\usepackage{bbm}
\usepackage[dvips,letterpaper,text={6.5in,9in}]{geometry}

\usepackage[english]{babel}
\usepackage{hyperref}
\hypersetup{
    colorlinks=true,
    linkcolor=blue,
    filecolor=magenta,      
    urlcolor=cyan,
}
\usepackage{tabu, multirow,color,dcolumn,bm,enumerate}

\usepackage{graphicx,bm}

\newcommand{\beq}{\begin{equation}}
\newcommand{\bea}{\begin{eqnarray}}
\newcommand{\eeq}{\end{equation}}
\newcommand{\eea}{\end{eqnarray}}
\newcommand{\bal}{\begin{align}}
\newcommand{\eal}{\end{align}}

\usepackage{tikz}
\usetikzlibrary{decorations.pathmorphing,decorations.markings}
\tikzset{
photon/.style={decorate, decoration={snake,amplitude=4pt, segment length=7pt}, draw=black},
particle/.style={draw=black, postaction={decorate}, decoration={markings,mark=at position .5 with {\arrow[draw=black]{>}}}},
antiparticle/.style={draw=black, postaction={decorate}, decoration={markings,mark=at position .5 with {\arrow[draw=black]{<}}}},
gluon/.style={decorate, draw=black, decoration={coil,amplitude=3pt, segment length=4pt}},
higgs/.style={draw=black,dashed,thick },
arrow/.style={draw=black, very thick, postaction={decorate}, decoration={markings,mark=at position 1 with {\arrow[draw=black]{>}}}}
}

\usepackage{latexsym}
\usepackage{subcaption}

\newcommand{\gev}{~\mathrm{GeV}}

\newcommand{\fb}{~\mathrm{fb}}

\newcommand{\ab}{~\mathrm{ab}}

\newcommand{\mz}{m_{Z}}
\newcommand{\mx}{m_{X}}
\newcommand{\gx}{g_{X}}
\newcommand{\sx}{s_{X}}

\newcommand{\btau}{b \bar b \tau^+ \tau^-}
\newcommand{\bl}{b \bar b \ell^+ \ell^-}
\newcommand{\ttbar}{b \bar b V_{\ell+\tau_\ell}V_{\ell+\tau_\ell}}

\usepackage{outlines}
\usepackage{enumitem}
\setenumerate[1]{label=\Roman*.}
\setenumerate[2]{label=\Alph*.}
\setenumerate[3]{label=\roman*.}
\setenumerate[4]{label=\alph*.}

\newcommand{\MET}{$E_T{\hspace{-0.47cm}/}\hspace{0.35cm}$}

\newcommand{\Lc}{\mathcal{L}}

\newcommand{\G}{\mathcal{G}}

\newcommand{\U}{U(1)_{B-L}^{(3)}}

\definecolor{darklightsabergreen}{rgb}{0.0, .49, 0.06}

\newcommand{\met}{\slashed{E}_T}

\begin{document}

\title{ LHC Constraints on a $(B-L)_3$ Gauge Boson}

\author{Fatemeh Elahi}
\affiliation{School of Particles and Accelerators, Institute for Research in Fundamental Sciences IPM, Tehran, Iran}
\author{Adam Martin}
\affiliation{Department of Physics, 225 Nieuwland Science Hall, University of Notre Dame, Notre Dame, IN 46556, USA}

\vspace*{0.5cm}
\begin{abstract}
\vspace*{0.5cm}

{
In this paper, we explore the constraints that the LHC can place on a massive gauge boson $X$ that predominantly couples to the third generation of fermions. Such a gauge boson arises in scenarios where the $B-L$ of the third generation is gauged. We focus on the mass range $10 \leq \mx \lesssim 2\,m_W$, where current constraints are lacking, and develop a dedicated search strategy. For this mass range, we show that $b \bar b \tau^+ \tau^-$, where at least one of the $\tau$s decay leptonically is the optimal channel to look for the $X$ at the LHC. The QCD production of $b$ quarks, combined with the cleanliness of the leptons coming from the decay of the $\tau$ allow us to detect  $X$ gauge boson with couplings of $\gx \sim(0.005-0.01)$, for $\mx < 50 \gev$, and a coupling of $O(0.1)$ for heavier $X$ gauge boson with $100 \fb^{-1}$ of integrated luminosity. This is about a factor of 2-10 improvement over previous constraints coming from the decay of $\Upsilon \to \tau^+ \tau^-$. Extrapolating to the full HL-LHC luminosity of $3000 \fb^{-1}$, the bounds on $\gx$ can be enhanced by another factor of $\sqrt{2}$ for $\mx < 50 \gev$. 
} 
\end{abstract}
\maketitle


\section{ Introduction}
\label{sec:Intro}

Even though numerous experimental measurements attest to the validity of the Standard Model (SM), some enigmatic observations such as dark matter and neutrino masses compel us to look for new physics (NP) beyond the SM. Many NP models propose augmenting the SM gauge groups by a new  gauge symmetry, with $U(1)$ being a popular choice. On the other hand, it is well-known that the SM Lagrangian respects some global $U(1)$ symmetries that are not demanded beforehand. Some of these so called ``accidental'' symmetrie are anomaly free and can be gauged, either within the SM alone or with minimal extension~\cite{Foot:1990mn, He:1990pn, He:1991qd, Foot:1994vd}. Given that the nature already approves of the SM, it is worthwhile to explore extending the SM using its own suggested symmetries. 

Among the possible symmetries, new interactions that involve electrons or the first two generation of quarks are severely constrained in various collider searches~\cite{Pati:1974yy,Marshak:1979fm,Wilczek:1979et,Mohapatra:1980qe, Nelson:2007yq,Harnik:2012ni} and low energy experiments~\cite{ delAmoSanchez:2010bt,Babu:2009nn, Babu:1999me,Golowich:2009ii}. The $U(1)_{L_\mu-L_\tau}$ symmetry -- the difference between muon and tau number -- has also received a lot of attention in recent years ~\cite{He:1991qd, Baek:2001kca,Ma:2001md,Salvioni:2009jp,Heeck:2011wj,Harigaya:2013twa,Carone:2013uh,Altmannshofer:2014cfa,Farzan:2015doa,Farzan:2015hkd,Elahi:2017ppe,Elahi:2015vzh,Chun:2018ibr}, and much of its parameter space is already being probed. That leaves us with new gauge bosons that interact predominantly with the third generation of fermions. One such possibility is $\U$, the difference between baryon number and lepton number of the third generation, which is anomaly free and guageable provided we augment the SM by a right handed neutrino.  

The $\U$ extension of the SM was first proposed by~\cite{1705.01822} to explain the flavor alignment of the third generation of quarks -- the empirical observations that the mixings of the third generation of quarks with the other two generations are very small.  Distinguishing the third generation by assigning it new quantum numbers under an additional symmetry prohibits the mixing with the third generation of quarks, and thus justifies its flavor alignment. Of course, the symmetry needs to be broken at some scale to allow small, yet non-zero mixing between the generations~\cite{Dev:2018pjn}.   

To achieve non-zero mixing between the generations at low scales, the $\U$ symmetry needs to be spontaneously broken by a scalar $\phi$ that is charged under $\U \times \G_{SM}$ and acquires a vacuum expectation value (vev). For certain charge assignments and coupling structure, it is possible to generate a realistic CKM matrix while relegating all tree-level flavor-changing-neutral currents (FCNC) to the up-quark sector~\cite{1705.01822} (see also Appendix~\ref{app:interaction}). The up-sector FCNC are suppressed by powers of CKM elements, however they -- along with the down-sector FCNC they generate at loop level -- are constrained by multiple low energy experiments. The constraints from experiments such as BaBar~\cite{delAmoSanchez:2010bt}, E949~\cite{Anisimovsky:2004hr,Artamonov:2008qb},  BESIII~\cite{Ablikim:2014uzh}, and CHARM-II~\cite{Vilain:1994qy} are severe, but peter out once $\mx \gtrsim 5\, \gev $. Furthermore, the direct coupling of $X$ with third generation of fermions can also contribute to the decay of $ \Upsilon \to \tau \tau$, which constrains the available parameter space for $X$ near $\mx \sim m_\Upsilon \simeq 10 \gev$~\cite{1002.4358}, however the contribution of off-shell $X$ to $\Upsilon$ decay dies off rapidly as we move away from $m_\Upsilon$. 

The $\mx$ window $\gtrsim 10\,\gev$ is only loosely constrained and is therefore the focus of this study. In practice, we impose an upper limit of $\mx < 2m_W$, as the interaction of $X$ with $W$ gauge bosons is closely tied to the mixing angle between Higgs and the $\U$ breaking scalars $\phi$, and thus introduces multiple additional parameters; $X$ in this range could conceivably be constrained by LHC resonant diboson searches such as Ref.~\cite{Aaboud:2017fgj,Sirunyan:2017acf,Sirunyan:2017nrt,CMS:2017skt,Biesuz:2017zip}. For even larger $\mx > 2 m_t$, $X$ phenomenology is driven by decays to top pairs. In this sense, $X$ phenomenology can be mapped into $Z' \to t\bar t$ searches, which have been studied extensively~\cite{Hill:1991at,Hill:1993hs,Hill:1994hp,Harris:2011ez,Rosner:1996eb,Lynch:2000md,Carena:2004xs,Choudhury:2007ux,Khachatryan:2015sma,CMS:2018ohu,Aaboud:2018mjh,Sirunyan:2017uhk,Sirunyan:2017yar,Cerrito:2016qig,Arina:2016cqj,Pedersen:2015knf,Fox:2018ldq}.

Having selected the $X$ mass window we are interested in, the next step is to determine the optimal LHC $X$ production mode and decay channel. As $X$ has suppressed couplings to first and second generation fermions, we either have to rely on the $b$ parton distribution function (PDF) (for $pp \to X$), or to produce the $X$ in association with third generation fermions, e.g. $pp \to \bar f f X$ where $f = t/b/\tau/\nu_{\tau}$. The PDF of b-quark is small, therefore we focus on associated production\footnote{While smaller, we do include processes initiated by $b$ PDFs in all our analyses.}.  The production of colored objects at the LHC is significantly larger than leptons, therefore we will concentrate on the scenario where $X$ is produced in association with a pair of $b$ quarks. Associated production of $X$ with top quarks is also an option, but suffers in rate due to the increased energy requirement as well as in reconstruction complexity, so we do not consider it here. 
 
Turning to $X$ decay, if $X$ decays to a pair of $b$ quarks, we have a four b final state, which makes QCD backgrounds overwhelming and introduces a combinatorics problem. Among the leptonic decays of $X$, $\tau$s are more preferable because they give more handles for kinematic variables. Thereby, we settle on $p p \to  b \bar b X \to b \bar b \tau^+ \tau^-$.  

One may think that further focusing the search on the $Z$ resonance contribution is a useful way to suppress backgrounds, as done in Ref.~\cite{Elahi:2015vzh} for the case of  $U(1)_{L_\mu-L_\tau}$ gauge bosons in $pp \to 4\,\mu$. However, the poorer energy resolution for jets (as compared to muons in Ref.~\cite{Elahi:2015vzh}) and the inevitable missing energy from neutrinos in tau decay hamper this technique and we find it is more beneficial to focus on QCD-produced $b\bar b$ pairs that emit an $X \to \tau^+\tau^-$.

The channel $ b \bar b \tau^+ \tau^-$ has already attracted some attention at the LHC in the search for the third generation leptoquarks ~\cite{CMS:2018pab,Khachatryan:2014ura} and di-Higgs searches~\cite{Sirunyan:2017tqo,Sirunyan:2018two,Cadamuro:2017wma}. However, due to their particular optimized cuts, these analyses will have limited-to-no sensitivity to $X$ in our mass range of interest. More specifically, 
\begin{itemize}
\item The search for the third generation leptoquarks ~\cite{CMS:2018pab,Khachatryan:2014ura} is ineffective because they impose $m_{\tau b} > 250 \gev$, whereas we find that our signal prefers $m_{\tau b} < 150 \gev$ for $\mx \le 2\, m_W$.
\item the CMS di-Higgs search~\cite{Sirunyan:2017tqo} considers $ b \bar b \tau^+ \tau^-$  in the mass window $400 \gev < m_{bb\tau \tau } < 700 \gev$. In our signal, however, the production of $b\bar b X$ is maximum at threshold, which means even for $\mx =2\,m_W$, we expect most of our events to lie in the $ m_{bb\tau \tau } < 350 \gev$ region. 
\item the results of other CMS di-Higgs searches~\cite{Sirunyan:2018two,Cadamuro:2017wma} are not easily recastable because they use boosted-decision-tree (BDT).
\end{itemize}  

Given the lack of constraints from the current LHC searches or any other experiments, in the following sections we develop a LHC search strategy for $\U$ gauge bosons, $10\,\gev \le \mx \le 2\,m_W$, using the $b \bar b \tau^+ \tau^-$ final state. We will assume throughout that $X$ is short-lived and therefore focus on prompt signals. Long-lived $X$, leading to displaced vertices at the LHC may be interesting to study, but likely require extending the setup in some way\footnote{In the current setup, the $X$ lifetime and production rate are governed by the same coupling, so one cannot make the particle long-lived without killing the production rate.}. For the case of fully hadronic $b \bar b \tau^+ \tau^-$ (prompt), the QCD backgrounds are overwhelming. Therefore, we will narrow our attention to semi-leptonic and fully leptonic decays of $\tau$s. Despite the large SM backgrounds (e.g., $t \bar t \to b \bar b W^\pm W^\pm \to b\bar b \tau^+ \tau^- + \met$), we show that the LHC-13 TeV, with the currently luminosity, can significantly improve the bounds on the $\U$ gauge coupling $\gx$.

The organization of the rest of the paper is as follows. In the upcoming section (Sec.~\ref{sec:model}), we introduce the model, including the free parameters we will consider for the phenomenology of $X$ gauge boson at the LHC. Next, in Sec.~\ref{sec:analysis}, we explore the LHC power in improving the bounds using simple kinematic variables -- both for $\mx < \mz$~(Sec.~\ref{sec:light}) and for slightly heavier $\mx \gtrsim \mz$~(Sec.~\ref{sec:heavy}). Finally, some concluding remarks are presented in Sec.~\ref{conclusion}.

\section{The $\U$ model }
\label{sec:model}

We study a model where the SM gauge symmetries are extended to include $\U$ symmetry -- the difference between the baryon number and the lepton number of the third generation. This symmetry is anomaly free, provided that we include a right-handed tau neutrino $\nu_{3R}$ to the SM. The charge assignments of the fermions are: $(Q_{3L}, u_{3R}, d_{3R}) : 1/3$ and $(\ell_{3L}, e_{eR}, \nu_{3R}): -1$, with all first and second generation fermions inert. 

From various observations, we know the exact $\U$ symmetry is not realized in nature at low scales, and thus must be broken. The simplest mechanism to spontaneously break $\U$ is to add some scalars charged under $\U$ symmetry that acquire vacuum expectations values (vev). To make the model phenomenologically viable, we actually have to introduce two $\U$-charged scalars, a SM singlet $s$ with  $\U$ charge $+1/3$ and $\phi$, a SM $SU(2)_\text{W}$ doublet with hypercharge $+1/2$ (identical SM charges as the Higgs) and $\U$ charge $+1/3$~\cite{1705.01822}. The Table of particles charged under $\U$ is shown in Table~\ref{tab:charge}. 
\begin{table}[h!]
  \begin{center}
    \begin{tabular}{ | c c c  c c | } \hline
  && $SU(3)_c \times SU(2)_\text{W} \times U(1)_{\rm Y}$ &&$ \U $\\ 
  \hline \hline
  $\phi $&&$({\bf 1}, {\bf 2}, 1/2) $&&$1/3$\\
    $s $&&$({\bf 1}, {\bf 1}, 0) $&&$1/3$\\
     \hline
  $Q_{3L} $&&$({\bf 3}, {\bf 2}, -1/6) $&&$1/3$\\
  $t_R $&&$({\bf 3}, {\bf 1}, 2/3) $&&$1/3$\\
  $b_R $&&$({\bf 3}, {\bf 1}, -1/3) $&&$1/3$\\
  $L_{3L} $&&$({\bf 1}, {\bf 2}, -1/2) $&&$-1$\\
  $\tau_R $&&$({\bf 1}, {\bf 1}, -1) $&&$-1$\\
   $\nu_{\tau R} $&&$({\bf 1}, {\bf 1}, 0) $&&$-1$\\
   \hline
    \end{tabular}
  \end{center}	
  \caption{Scalar and fermion fields charged under the $\U$ gauge symmetry.}
\label{tab:charge}
\end{table}
\noindent The $\phi$ field is needed to connect first and second generation quarks to the third generation quarks via renormalizable interactions, while the additional source of $\U$ breaking from the $s$ field allows us to decouple the mass of the  $\U$ gauge boson $X$ from the electroweak breaking scale. Note that Yukawa terms involving only third generation fields involve the Higgs, not $s$ or $\phi$, and that renormalizable inter-generation interactions involving the third generation
between leptons are forbidden by the $\U$ charge assignment. Neutrino masses can be accommodated via higher dimensional operators or via further extensions of the model by vector-like matter~\cite{1705.01822}. 

In this paper, we are interested in the phenomenology of the $X$ gauge boson. The $X$ gauge boson appears in the covariant derivative of the third generation fermions, indicating a tree-level interaction of $X$ with third generation of fermions in the interaction basis. Another place $X$ appears is the covariant derivative of scalars ($s$ and $\phi$), which not only results in $X$ acquiring a mass (once $\langle \phi \rangle, \langle s \rangle \ne 0$), but also leads to tree-level interactions of $X$ with scalars. Furthermore, because $\phi$ is charged under both $SU(2)_\text{W} \times U(1)_\text{Y}$ and $\U$, its kinetic term induces a mixing with $X$ and $Z$ gauge boson with an angle 
\beq
\sx \equiv \frac{2}{3} \frac{\gx}{\sqrt{g^2+g^{'2}}} \frac{v_\phi^2}{v^2},
\eeq
where $\gx$ is the gauge coupling associated with $\U$, $(v_\phi)$ represents $\phi$ vev, and  $v =\sqrt{ v_\phi^2 + v_h^2} = 246 \gev$, with $v_h$ being the Higgs vev. Therefore, in the mass basis, the  (mass eigenstate) $X$ boson interacts with $Z$ current with a coupling proportional to $\sx$, while the (mass eigenstate) $Z$ boson interactions will be modified by an amount proportional to $\sx$. 

In addition to $X$, the model contains several new scalars (from $\phi, s$) and a right-handed neutrino. For simplicity, and following Ref.~\cite{1705.01822}, we assume that these states are all heavier than $\mx/2$ so they play no role in our analysis.  

 The relevant model parameters to study $X_\mu$ phenomenology are the $X$ mass  ($\mx$), the $\U$ gauge coupling ($\gx$), and the rotation angle between $Z$ and $X$ ($\sx$). Rather than use $\sx$, we find it more convenient to work with $\tan{\beta} = v_h/v_\phi$. In terms of these parameters, 
 \beq 
\mx ^2 = \frac{1}{9} \gx^2 \left(\frac{v_\phi^2 v_h^2}{v^2} + v_s^2\right) = \frac{1}{9} \gx^2 \left( v^2 \frac{ \tan^2 \beta }{(1+ \tan^2 \beta)^2} + v_s^2\right).
\eeq 
Notice that the presence of $v_s$ means $\mx$ is not tied to the electroweak scale and can, in principle, be large.
 
In the gauge interaction basis, the interaction between fermions and  the (mass eigenstate) $X$ gauge boson has the form $ c_\alpha\,\bar f_{\alpha} \gamma^{\mu} f_{\alpha} X^\mu,$ with
\beq
 c_\alpha = \gx q_\alpha^X + \sx \sqrt{g^2 + g^{'2}} q_\alpha^Z =\gx \left[q_\alpha^X + \frac{2}{3}  q_\alpha^Z (1+ \tan{\beta}^2)^{-1}\right].
\eeq 
Here, $q_\alpha^X$ and $q_\alpha^Z= I_3^\alpha - s_w^2 q_\alpha$ are respectively the $X$ and $Z$ charge of fermion $\alpha$, $q_\alpha$ is the electric charge, $s_w$ is the $\sin \theta_{\text{weak}}$, and $I_3^\alpha$ is the $W^3_\mu$ generator. The translation of this interaction to the fermion mass basis induces flavor changing interactions among left handed up-type quarks and is shown in detail in Appendix~\ref{app:interaction}. 

An important property of $X$ for our study is how it decays to various SM states. Due to $X-Z$ mixing, the branching ratio of $X$ strongly depends on the value of $\tan{\beta}$. For small $\tan{\beta}$, the coupling of $X$ to the $Z$ current is important, while for large $\tan{\beta}$ $X$ predominantly decays to third generation fermions. The branching ratio of $X$ to various SM final states for $\tan{\beta} = 1$ and $\tan{\beta} = 5$ is shown below in Fig.~\ref{fig:branching}~\footnote{There is a constraint on the value of $\tan{\beta}$ coming from Higgs coupling measurements, roughly between $1\, \le \tan{\beta} \le 30$. The exact range depends on the scalar $h, \phi, s$ spectrum, the details of which we ignore here, therefore we will work with this approximate range. }. In this figure, we derived the branching ratio of $X$ to hadrons using Ref.~\cite{Whalley:2003}, and we have assumed the new scalars and the sterile neutrino are heavier than $\mx/2$. We can see that the branching ratio of $X$ to a pair of $\tau$s dominates for $\mx \gtrsim 5\, \gev$ (and up to $\mx \sim 2m_W$). This channel dominates because of the relatively large values of $q_\tau^X$ and $q_\tau^Z$ compared to other third generation fermions. 

 \begin{figure}[t!]
 \centering
 \includegraphics[width=.5\textwidth]{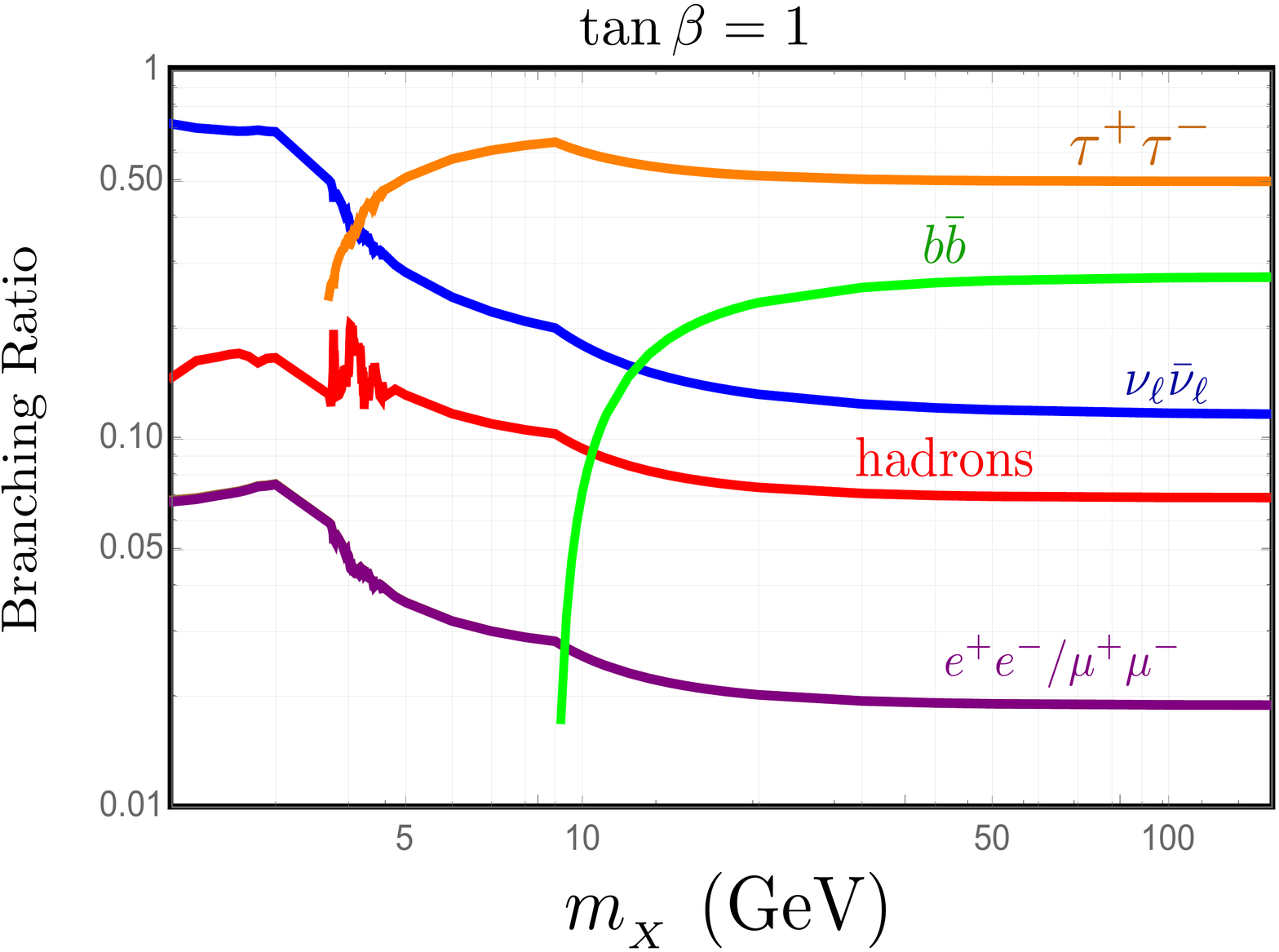}~
 \includegraphics[width=.5\textwidth]{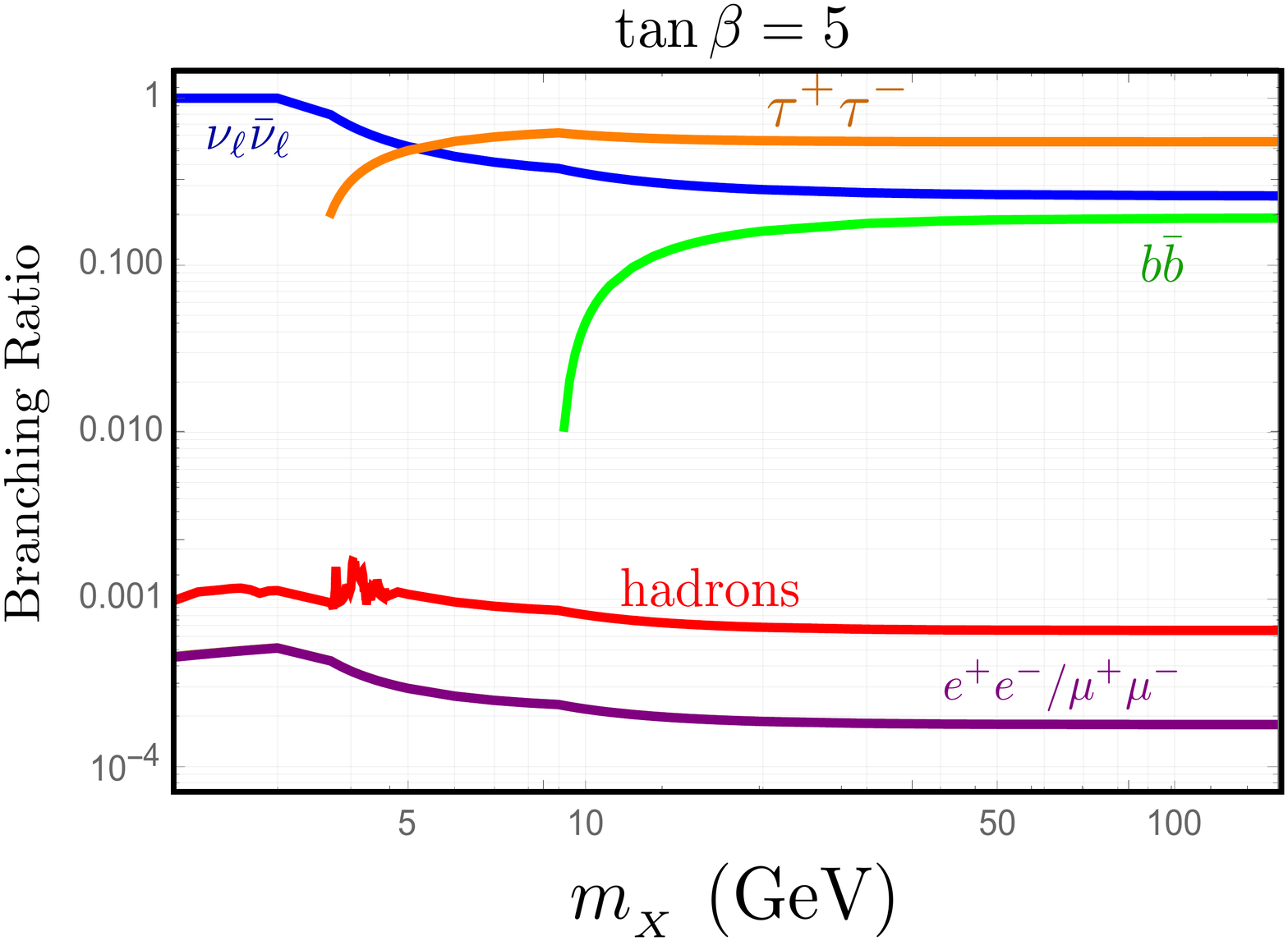}
  \caption{The branching ratio of $X$ to various final states for $\tan{\beta} = 1$ ({\bf Left}) and $\tan{\beta} = 5$ ({\bf Right}), with the assumption that the new scalars and the sterile neutrino are heavier than $\mx/2$. The branching ratio differs significantly depending on the value of $\tan{\beta}$. The highest branching ratio in the heavy mass region is to $\tau^+ \tau^-$, due to a combination of large $q_{_X}$ and large EM charge.} 
 \label{fig:branching}
\end{figure}

Having defined the model, we now move on to its LHC signatures. As mentioned earlier, our focus is on $\mx > 10 \gev$ where low-energy constraints are absent. For $\mx > 10 \gev$ the only non-LHC bounds are from $\Upsilon \to \tau^+ \tau^-$ and the modification of the oblique parameters. The contribution of off-shell $X$ to $\Upsilon$ decay dies off rapidly as $1/\mx^2$, and the constraints coming from oblique parameters become very mild for $\tan \beta > 2$.

\section{ LHC Constraints }
\label{sec:analysis}

The main advantage of the LHC is that it can produce the $X$ gauge boson on-shell. This fact is crucial, because amplitudes containing off-shell $X$ are suppressed by two powers of $\gx$ -- one at the production vertex and another at its destruction. On-shell exchange, on the other hand, comes with only one factor of $\gx$ at the $X$ production vertex (amplitude level) as the decay portion contributes some $O(1)$ branching ratio factor. 

The chief way to produce an on-shell $X$ at the LHC is in associated production with a pair of $b$ jets, $p p \to b \bar b X$. 
 In such processes, we can benefit from the large QCD production of $b$s as well as the sizable coupling of $X$ with $b$ quarks. The $X$ boson can decay in many ways, however, we will focus here on $p p \to b \bar b X \to b \bar b \tau^+ \tau^-$.  This choice is motivated by the large $\text{Br}(X \to \tau^+ \tau^-)$, however there are some other  important benefits:
 \begin{itemize}
 \item large number of observables to help signal background discrimination, in contrast to $b\bar b + \met$ production. 
\item there are no combinatorics issues, as opposed to the $\bar b b \bar b b$ final state. 
\end{itemize}
 
Because $\tau$s are not stable at the LHC, the search mode has to be further defined in terms of the $\tau^+ \tau^-$ final decay products. 
While there exists several options, we find that requiring at least one of the $\tau$s to decay leptonically is necessary to suppress the (otherwise enormous) QCD background.   
To decide between semi-leptonic $\tau$s or fully leptonic ones, let us turn our attention to potential triggers\footnote{As we will show in the subsequent sections, the \MET distribution in the signal favors lower values. Therefore, a \MET trigger is also not ideal for our study}. As leptons are relatively  clean objects at the LHC, they have softer trigger cuts, and the presence of multiple leptons softens the requirement on each lepton further. As an example, the single lepton trigger at CMS ~\cite{Khachatryan:2016fll,1611.04040} requires $p_T (\ell) > 23 \ (20) \gev$ if the lepton is electron (muon), and the CMS di-lepton trigger~\cite{Khachatryan:2016fll,1611.04040} requires $p_T(\ell_1) > 17 \gev$ and $p_T(\ell_2) > 12\ (8) \gev$ with the second leading lepton being electron (muon)\footnote{The ATLAS numbers trigger cuts are similar: single lepton requires the $p_T(\ell)  > 26 \gev$~\cite{Aaboud:2017ojs,Aaboud:2017buh}, while the di-lepton trigger requires $p_T (\ell) > 17 \gev$, on both of the leading leptons~\cite{Aaboud:2017buh}.}.
Production of $b\bar b X$ is dominated near threshold (rather than with boosted $X$), hence the leptons coming from a light $X$ are expected to have small $p_T$. Therefore, for low $\mx$ the fully leptonic $\tau$s is the better option since the softer thresholds in the dilepton trigger will accept more signal. For high $\mx$, the leptons from $X$ decay are significantly energetic to be picked up efficiently by the single lepton trigger. This makes the semi-leptonic mode viable, and its larger branching fraction (compared to dileptonic taus) partially compensates for the drop in the signal cross section as $\mx$ increases.

As the optimal $b\bar b \tau^+\tau^-$ final state depends on $\mx$, we divide our analysis to two sections. In the following section (Sec.~\ref{sec:light}), we study $X$ gauge bosons with $ \mx < \mz$ using the  $b\bar b  \ell^+ \ell^- \met$ final state. Then, in Sec.~\ref{sec:heavy}, we use the $b\bar b \ell j \met$ final state to explore heavier $X$ gauge bosons, $\mx \sim [\mz, 2\, m_W)$. To thoroughly study the LHC detection prospects, we generated a Universal FeynRules Output (UFO) model~\cite{Degrande:2011ua} using {\tt Feynrules}~\cite{Alloul:2013bka}. We then fed the model to {\tt MadGraph-aMC@NLO}~\cite{Alwall:2014hca,Alwall:2011uj} for all simulations\footnote{We used the nn23lo1 parton distribution functions for all event samples, with factorization scale and renormalization scale set to their default value, $\hat s$.}, including the calculation of $X$ total width for a given coupling ($\gx$ and $\sx$). We used {\tt Pythia 8.2} ~\cite{Sjostrand:2014zea} for hadronization, showering, and $\tau$ decay, and {\tt Delphes}~\cite{deFavereau:2013fsa}  with default cards for detector smearing, flavor tagging and jet reconstruction.

\subsection{ Light $X$: $\mx < \mz$ }
\label{sec:light}
\begin{figure}[t!]
 \includegraphics[width=.6\textwidth]{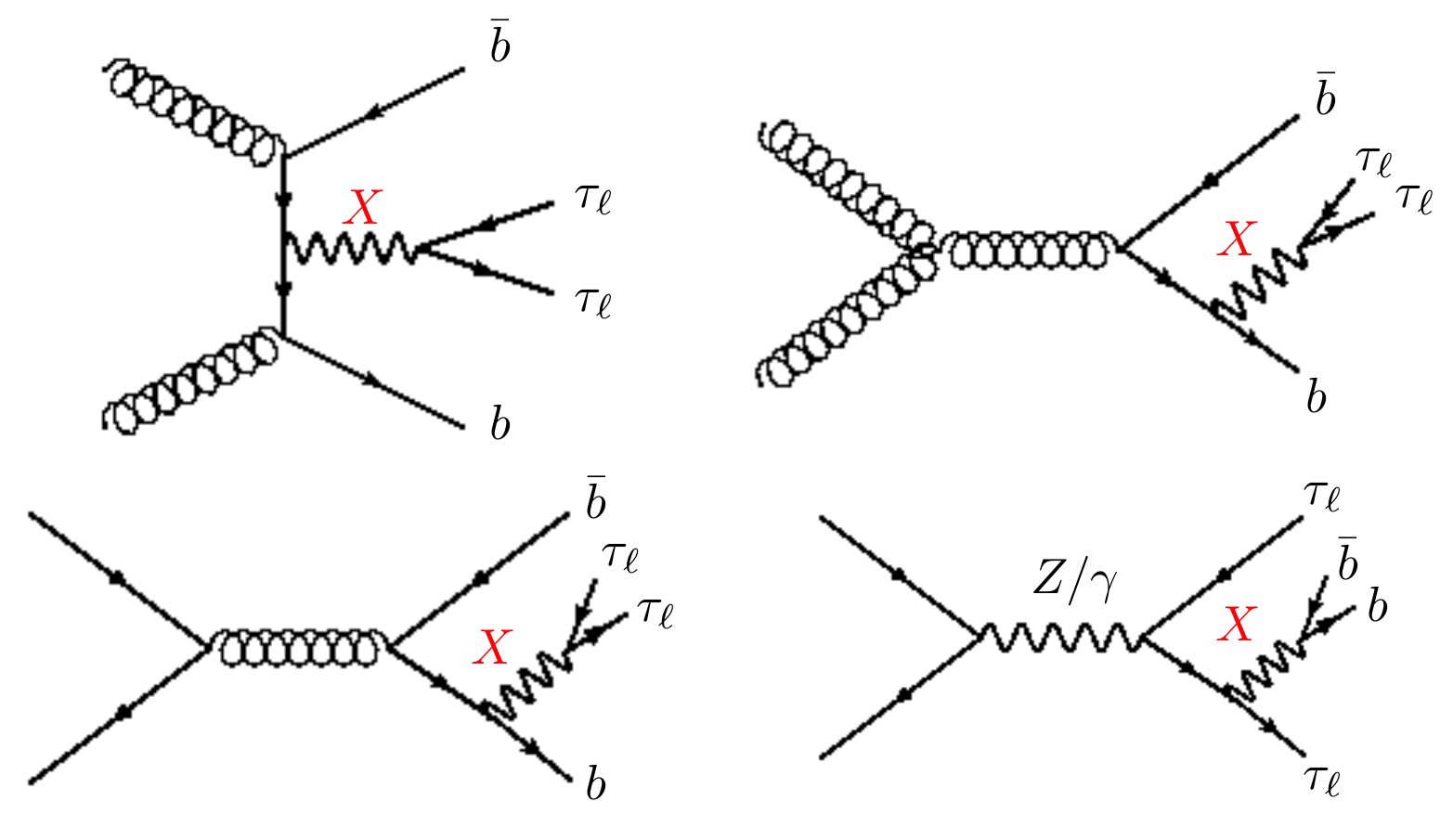}
 \caption{The signal Feynman diagrams corresponding to the production of $ b \bar b \ell^+ \ell^- \met$ final state. }
 \label{fig:feynsig}
\end{figure}

As discussed above, for $\mx < \mz$ the final state we are interested in extracting is $ b \bar b \ell^+ \ell^- \met$.   
The main contribution to the signal comes from $ p p \to b \bar b X$ with $X \to \tau^+ \tau^-$ (specifically, the process $ p p \to \tau^+ \tau^- X$ followed by $ X \to b \bar b $ only contributes to the signal at sub-percent level); Fig.~\ref{fig:feynsig} shows some of the signal processes. Because $X$ is predominantly produced on-shell, the signal has little interference with the SM backgrounds. Therefore, to a good approximation, the cross section can be expressed as: 
\beq
\sigma (p p \to b \bar b \tau^+ \tau^-)_{_{\text{Signal}}} \sim  c_b(\gx, \tan{\beta})^2 \times f(\mx, s) \times \text{Br} (X \to \tau^+ \tau^-)_{\tan{\beta}},
\label{eq:xsection}
\eeq 
where $c_b(\gx, \tan{\beta})$ is the coupling of $X$ with b quarks. The subscript $\tan{\beta}$ in Eq.~\eqref{eq:xsection}, indicates that the branching ratio of $X$ depends on $\tan \beta$. Technically, the branching ratio depends on $\mx$ as well, however, for the mass range of $2 m_b \leq \mx < 2 m_{_W}$ the branching ratios are constant with respect to $\mx$. The remaining part of the cross section, $f(\mx, s)$, governs the kinematics and is a function of $\mx$ and collider energy $s$ only. In our simulations, we generated events for $ \mx = 10,\  20,\  30, \ 50,$ and $80 \gev$, and fixed $\gx = 0.02$ and $\tan{\beta} = 5$. 
However, as $c_b(\gx, \tan{\beta})\times \text{Br} (X \to \tau^+ \tau^-)_{\tan{\beta}}$ do not play a noticeable role in the kinematic distributions (which govern signal acceptances), the LHC sensitivity at one $\gx, \tan{\beta}$ value can be extended to other  $\gx, \tan{\beta}$ values simply by rescaling:
 \beq
\sigma^{\text{New}}_{_{\text{Signal}}} \sim \sigma^{\text{Old}}_{_{\text{Signal}}} \left(\frac{ c_b(\gx^n, \tan{\beta}^n)}{c_b(\gx^o, \tan{\beta}^o) }\right)^2 \times \frac{\text{Br}(X\to \tau^+ \tau^-)_{\tan{\beta}^n}}{\text{Br}(X\to \tau^+ \tau^-)_{\tan{\beta}^o}},
\label{eq:gx}
  \eeq
 where the indices $n$ and $o$ refer to new and old, respectively. 
  
 There are a number of SM process that give rise to the $b \bar b \ell^+ \ell^- \met$ final state, namely:
\begin{eqnarray}
1.) \quad &&  pp\rightarrow\,  b \bar b V_{\ell+\tau_\ell} V_{\ell+\tau_\ell}: \left\{ \begin{array}{c}b \bar b W^+_{\ell+\tau_\ell}   W^-_{\ell+ \tau_\ell} \hspace{0.5 in} \\\\  b \bar b (Z/\gamma^{\star})_{\ell+\tau_\ell} Z_{\nu}\ \ \ \ \ \end{array} \right.\nonumber  \\
 2.) \quad  && pp\rightarrow\, b \bar b \tau_\ell^+ \tau^-_\ell \nonumber \\
3.) \quad && pp \rightarrow\, b \bar b \ell^+ \ell^- 
\label{eqn:backgrounds}
 \end{eqnarray}
where $ \ell = e, \mu$, and $(W/Z)_{\ell+ \tau}$ refers to all possible charged leptonic decay of $W/Z$. Similarly, $\tau_\ell$ refers to the decay of $\tau$ to lighter leptons. The main difference between these three backgrounds is the number of neutrinos. The first background has two or six neutrinos, depending on whether the gauge bosons decay to $\ell$ or $\tau_\ell$, respectively,  the second background has four, and the third does not have any neutrinos. However, due to pile-up, jet mis-measurement, and the leptonic decays of a charged $B$ mesons in the b-jets, a net \MET can be generated, making the last background worthy of mentioning.

 \begin{figure}[t!]
 \includegraphics[width=.8\textwidth]{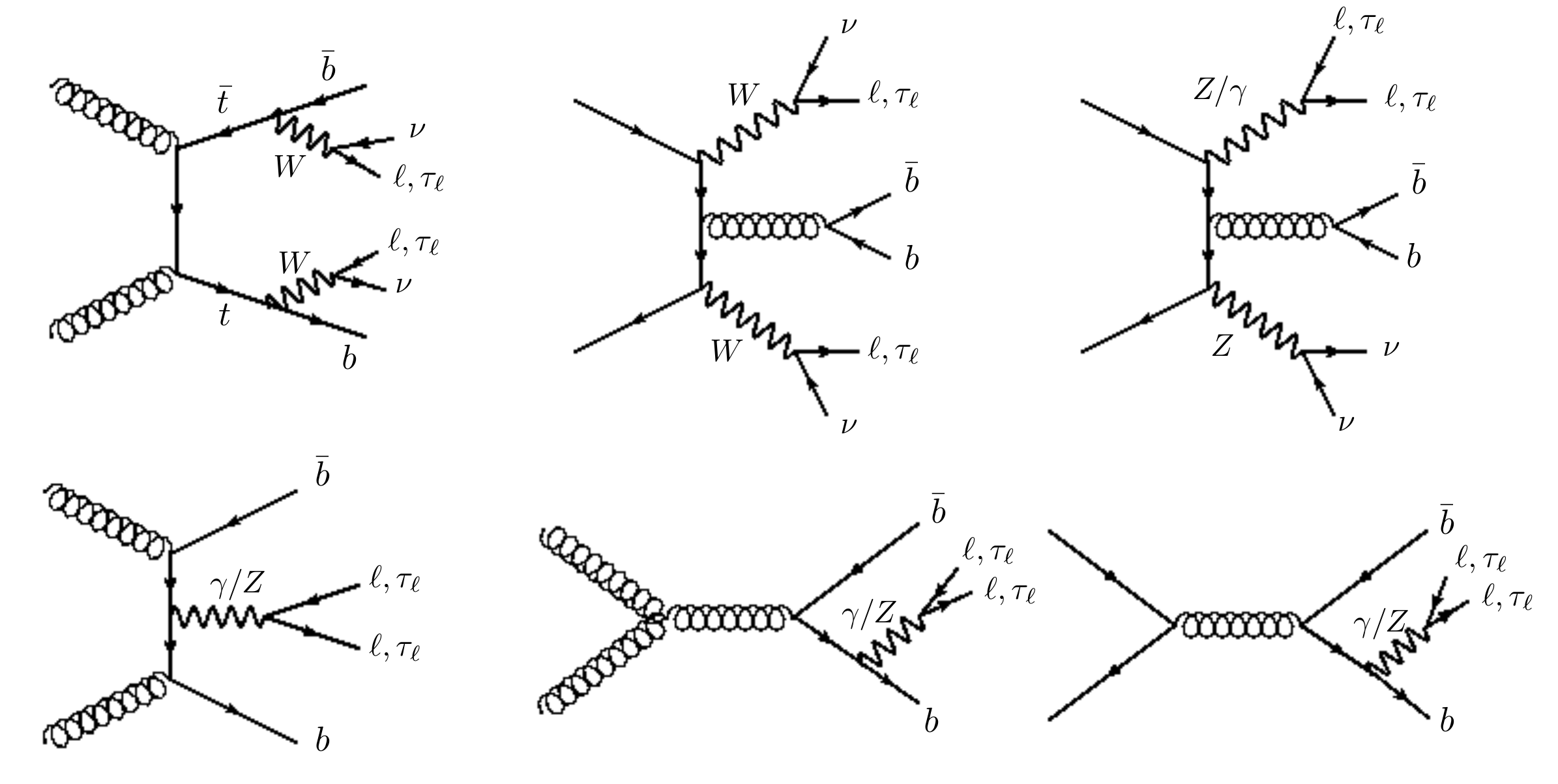}
 \caption{The SM backgrounds to $ b \bar b \ell^+ \ell^- + \met$ production at the LHC. In these diagrams $ \ell = \mu, e$, and $\tau_\ell$ refers to a $\tau$ that decays leptonically. The first line the Feynman diagrams corresponds to the first mentioned background in Eq.~\ref{eqn:backgrounds}. The second and third backgrounds in Eq.~\eqref{eqn:backgrounds} are shown in the second line of these Feynman diagrams, where the gauge bosons decay to $\tau_\ell$ and $\ell$, respectively.   
 }
 \label{fig:feynbkg1}
\end{figure}

 A few of the important Feynman diagrams for the SM production of $b \bar b \ell^+ \ell^- \met$ are shown in Fig.~\ref{fig:feynbkg1}. The largest irreducible background comes from $pp \to t \bar t \to b \bar b W^+ W^-$, as it is purely a QCD process with $\text{Br} (t \to b W) \simeq 1$. Other (continuum) processes contributing to the first background in Eq.~\eqref{eqn:backgrounds} do not have a large rate. 
 The second background, $ b \bar b \tau^+_\ell \tau^-_\ell$, is very similar to the signal, where $X$ is replaced with a $Z$ or $\gamma$.  Technically, the $\left(p p \to b \bar b \tau^+_\ell \tau^-_\ell\right)$ background and the signal can interfere, however the fact that $X$ are predominantly produced on-shell renders the interference is very small\footnote{ In our simulations, we force $X$ to be on-shell. To make sure this shortcut does not significantly influence our results, we tested the effects of off-shell $X$ (and interference) for various values of $\mx$. In all cases, the difference between on-shell $X$ and the full treatment was negligible.}. The last background is similar to the second background, but instead of producing a pair of $\tau$ leptons, $Z/\gamma \to \ell^+ \ell^-$. 
 
We must also include (reducible) backgrounds where a gluon/light flavor jet is mis-identified as a b-jet. The mis-identification rate of a $c$ jets is significantly higher than other light quark/gluon initiated jets (collectively referred to as $j_l$). Therefore, we considered charm-jet and light-jet processes separately. To include the impact of misidentifications, we add two versions of all backgrounds in Eq.~\eqref{eqn:backgrounds} -- one with $b$ jets replaced with $c$ and one with $b$ replaced by $j_l$. For example, the second background is expanded to include:
\begin{align}
2.) pp \to b \bar b\, \tau^+_{\ell} \tau^-_{\ell} \Rightarrow \left\{ \begin{array}{c} pp \to c\bar c \tau^+_{\ell} \tau^-_{\ell} \\ pp \to j_l\,j_l \tau^+_{\ell} \tau^-_{\ell} \end{array}\right.
\end{align}
Other backgrounds induced by lepton mis-identification are expected to have a very low rates~\cite{Khachatryan:2015hwa,Aaboud:2016vfy,Sirunyan:2018fpa,Aad:2016jkr} and thus are ignored in this study. The MC event samples for all processes are simulated at leading order (LO), with the overall rates scaled to next-to-leading-order (NLO)\footnote{Using MadGraph5-aMC@NLO for a 13 TeV LHC, we find the $\kappa_{\ttbar} \equiv\sigma_{\rm NLO}/\sigma_{\rm LO}$ of the $\ttbar$ process is roughly $1.4$, 1.8 for $\kappa_{\btau}$, 1.9 for  $\kappa_{\bl}$. For the reducible background where a light flavor/gluon jet is mis-identified as a b-jet, we assume $\kappa = 2$. The $\kappa$ of the signal is assumed to be $1.8$, due to the similar topology of the signal with the $\btau$ background. }. 

Before studying the kinematic distributions, we impose some preliminary cuts to ensure the events have been triggered upon, and that the visible final states are within the fiducial region of the detector. Specifically, we select events that satisfy the following requirements: 
 \begin{itemize}
 \item[] -- Include exactly two isolated opposite sign leptons (any combination of electrons and muons) with $|\eta|< 2.5 $ and separated from each other by $\Delta R > 0.4$. We further require the leptons to satisfy the dilepton trigger: the leading lepton must have $p_T(\ell_1) > 17 \, \gev$,  and the second leading lepton is required to have $ p_T(\ell_2) > 12 (8) \, \gev$ if the lepton is electron (muon).
 \item[] -- Include exactly two jets with $p_T(j) > 30 \gev$, $|\eta|< 2.5 $, and separated from each other by $\Delta R > 0.4$. Both jets must be $b$-tagged. We use the b-tagging option in {\tt Delphes}~\cite{deFavereau:2013fsa}, which corresponds to roughly a b-tagging efficiency of $60\%$, with a charm mistagging of $ 15\%$ and a light jet mis identification rate of $1\%$. 
 \end{itemize}
After these requirements, the largest background is $\ttbar$ with cross section of $1362 \, \fb$, followed by $\bl$ with $255\, \fb$ cross section, while that of $\btau$ is $28 \fb$. Among the reducible (fake $b$) backgrounds,  $c \bar c \ell^+ \ell^-$ ($20.3\, \fb$), $ j_l j_l \ell^+ \ell^-$ ($8\, \fb$), and $c \bar c \tau^+ \tau^-$ ($2.4\, \fb$)  are the most significant. The rest of the irreducible backgrounds have cross sections $\le 2 \, \fb$ (see Table~\ref{tab:eff}).

\begin{figure}[h!]
  \includegraphics[width=.49\textwidth]{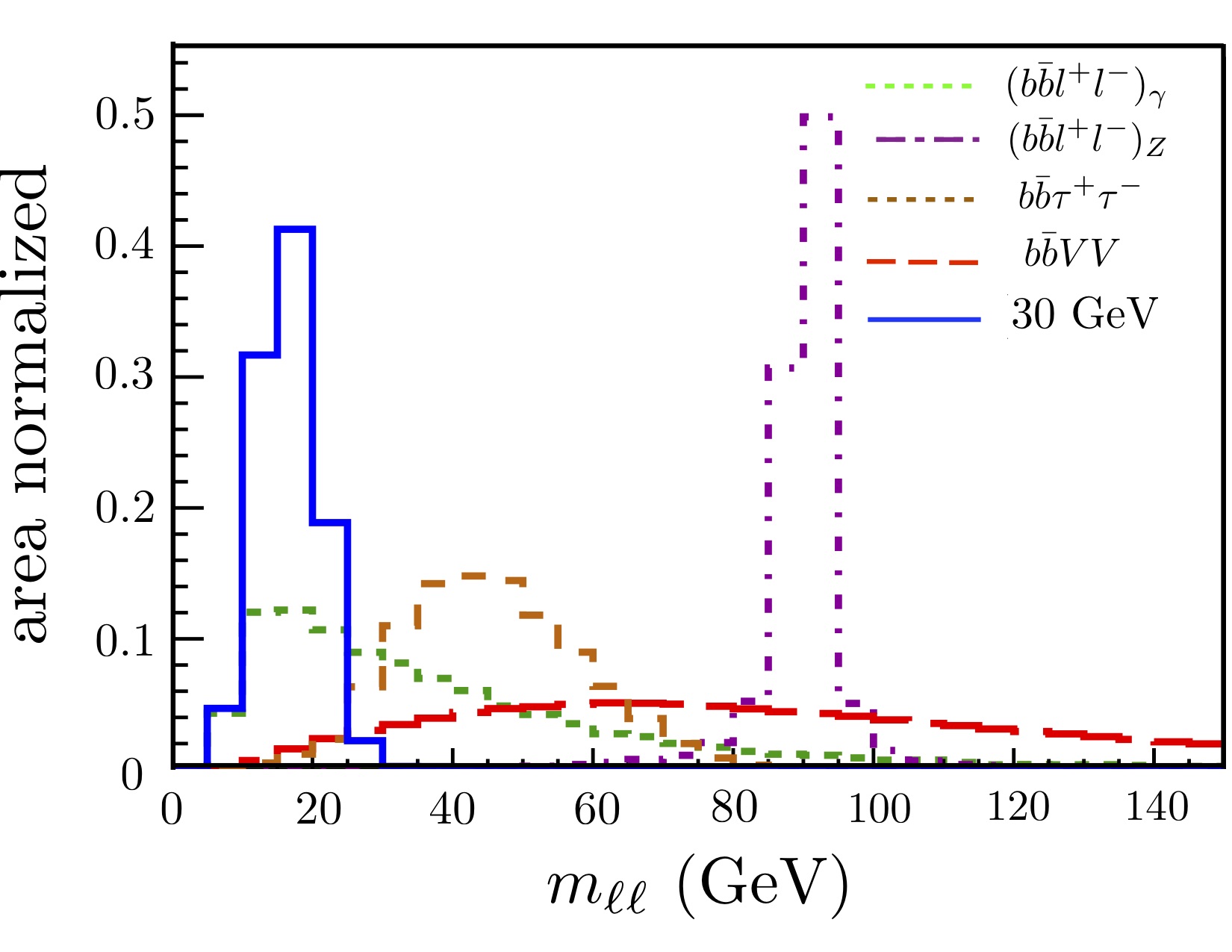}
 \includegraphics[width=.49\textwidth]{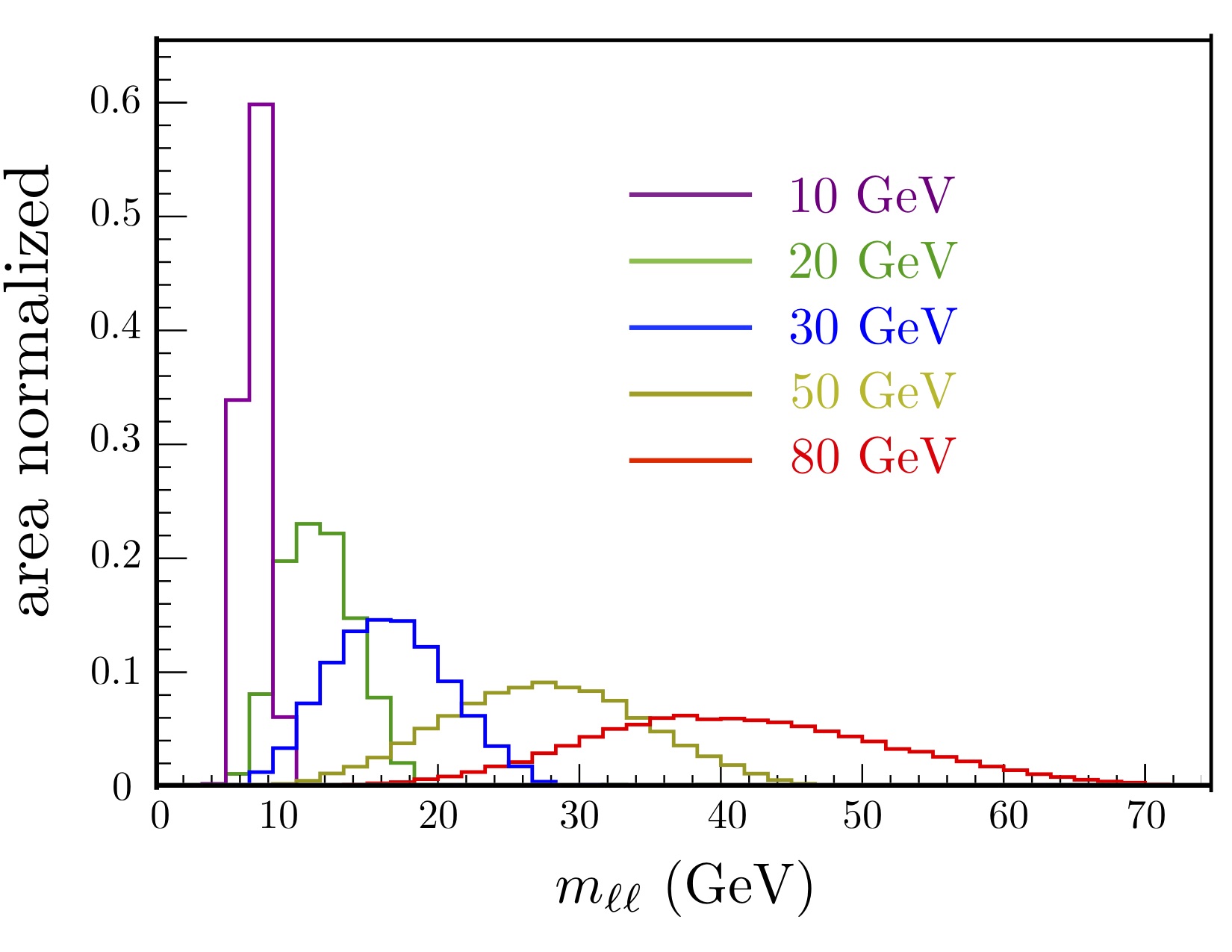}
 \caption{{\bf Left:} The area-normalized distribution of invariant mass of the two leptons ($m_{\ell \ell}$) for one of the signal benchmarks $\mx = 30 \gev$, compared with the dominant backgrounds. These distributions are after the basic cuts, and $ Z^{(\star)} \to \ell \ell$ and $\gamma^* \to \ell \ell$ have been listed as separate contributions since the interference between them is small. To make the distribution in less cluttered, we did not include the reducible backgrounds (Eq.~\ref{eqn:backgrounds2}), however the distributions of $c \bar c \ell^+ \ell^-$ and $ j_l j_l \ell^+ \ell^-$ behave like $\bl$, and  $c \bar c \tau^+_\ell \tau^-_\ell$ looks similar to $b \bar b \tau_\ell^+ \tau^-_\ell$. {\bf Right:}  The area-normalized $m_{\ell \ell}$ distribution for various $X$ masses are shown. The distribution of $m_{\ell\ell}$ becomes less faithful to $m_{_X}$ as we increase the mass of $X$.  }
 \label{fig:bench}
\end{figure}

 Thankfully, the topology of the dominant background -- $\ttbar$ -- is vastly different from the signal, giving us a hope to reduce it further with additional kinematic cuts. One variable that is particularly useful in teasing out the signal is the invariant mass of the leptons. In the signal we know $ m_{\ell \ell} < \mx$, while the invariant mass of the leptons in $\ttbar$ has a broad, featureless distribution. Therefore, requiring an upper bound on $m_{\ell \ell}$ can significantly suppress the $\ttbar$ background while retaining most of the signal region. A comparison of the $m_{\ell\ell}$ distribution (area normalized) for the background and a signal benchmark, $\mx = 30 \gev$ is shown below in the left panel of Fig.~\ref{fig:bench}.  
 
For light $\mx$, the only background that behaves similarly to the signal is $ pp \to \gamma^* \to \tau^+_\ell \tau^-_\ell, \ell^+\ell^-$. The cross section of this background is highly suppressed by the isolation cut $ \Delta R(\ell^+ , \ell^-) > 0.4$. The isolation cut also impacts the signal for small values of $\mx$,  however, for all benchmarks we are considering ($\mx \ge 10\, \gev$) imposing lepton isolation is more beneficial than relaxing it.

\begin{figure}[h!]
 \centering
  \includegraphics[width=.9\textwidth]{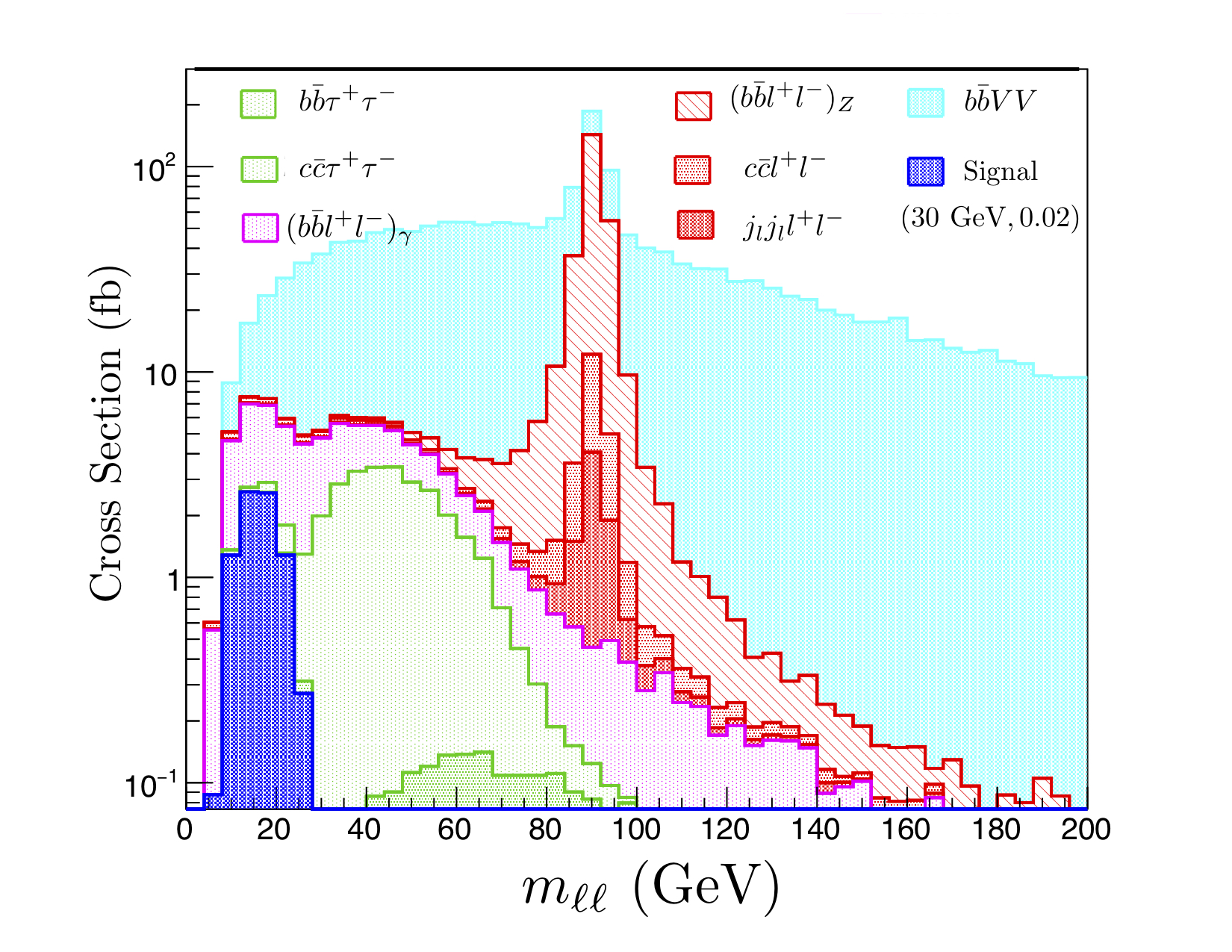}
  \caption{The $m_{\ell \ell }$ distribution of the backgrounds scaled according to their cross section. The signal benchmark chosen for this plot is $ (\mx, \gx) = (30 \gev, 0.02)$. However, since the cross section of the signal is very small compared to the backgrounds, we multiplied the signal distribution by a factor of 10 to make it more visible in the plot. This distribution is only after the basic cuts. By imposing $ m_{\ell \ell} < 25$, we can keep most of the signal, while throwing away a large portion of the background.}
 \label{fig:mllstack}
\end{figure}

To understand the discriminatory power of the $m_{\ell \ell}$ cut, we plot the $m_{\ell \ell}$ of backgrounds normalized to actual cross section, in Fig.~\ref{fig:mllstack}. The dark blue region is the distribution of the signal with $\mx = 30 \gev$. The cyan region belongs to the $\ttbar $ distribution, the red shaded region is the $b \bar b (\ell^+ \ell^-)_Z$, and the magenta dotted distribution belongs to $b \bar b (\ell^+ \ell^-)_\gamma$ distribution. The reducible background $c \bar c \ell^+ \ell^-$ and $j_l j_l \ell^+ \ell^-$ are also shown in red. The smooth green region belongs to $b \bar b \tau_\ell^+ \tau^-_\ell$, and the dotted green distribution is for $c \bar c \  \tau^+_\ell \tau^-_\ell$. Only basic selections have been imposed on these distributions. 

For our $\mx = 30 \gev$ benchmark, imposing $m_{\ell \ell} < 25 \gev $ eliminates 96\% of the background while retaining $97\%$ of the signal. For other $\mx$ benchmarks, an appropriately optimized $m_{\ell \ell}$ cut performs similarly, though its effectiveness decreases for larger $\mx$.  The decrease can be understood by looking at the $m_{\ell\ell}$ distribution for various $\mx$, shown in the right panel of Fig.~\ref{fig:bench}. As we increase $\mx$, the $m_{\ell \ell}$ distribution broadens and gets more separated from the $\mx$ value. The broadening occurs as a result of the allocation of the $X$'s energy between leptons and neutrinos. The power of $m_{\ell \ell}$ in distinguishing the signal for each of the benchmarks is tabulated in Appendix~\ref{app:cutflow}, where the upper bound on $m_{\ell \ell}$ (approximately $m_{\ell \ell} \lesssim 4/5 \mx$) has been optimized for each $\mx$ value.

\begin{figure}[t!]
 \centering
    \includegraphics[width=.5\textwidth]{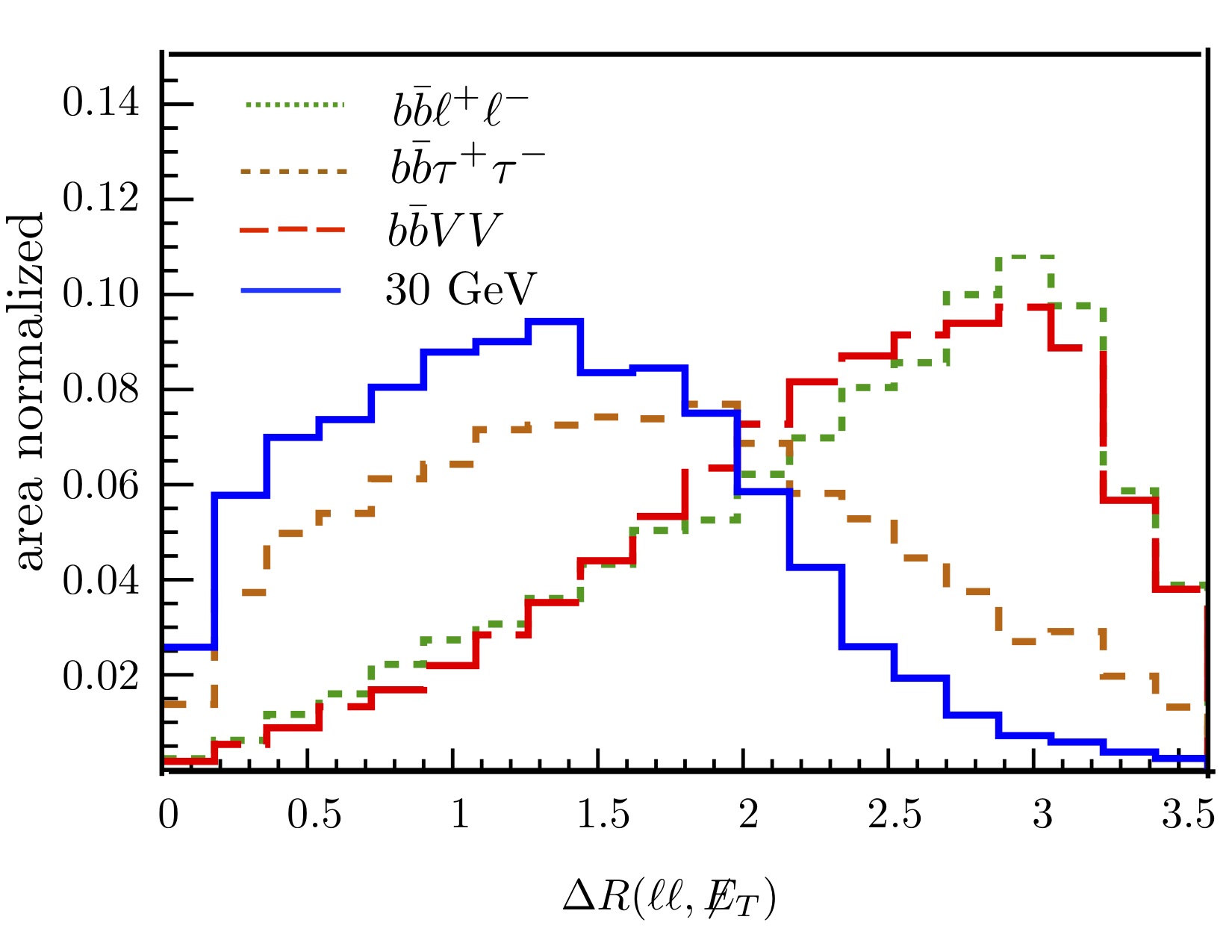}~~
      \includegraphics[width=.5\textwidth]{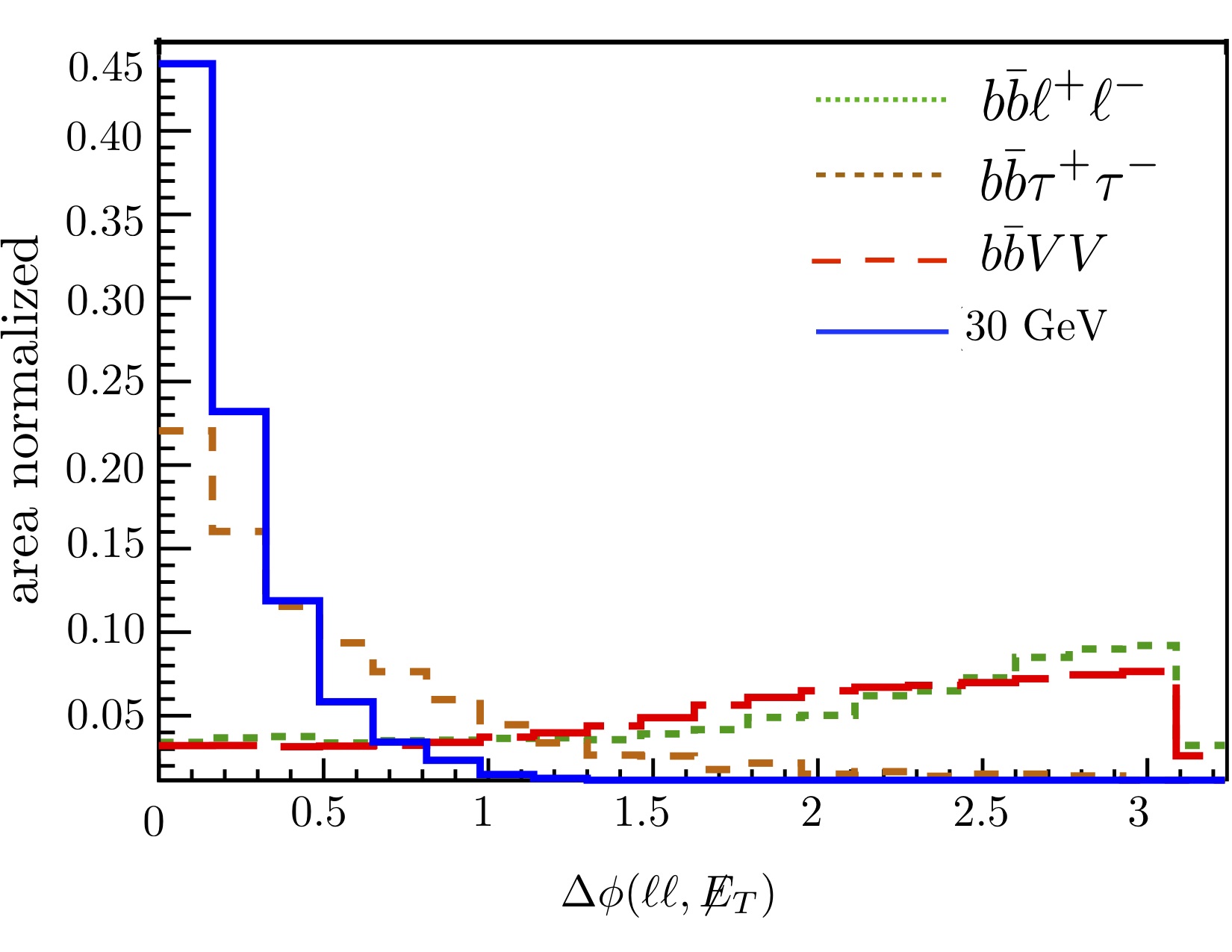}
  \caption{The comparison between area-normalized $\Delta R (ll, \met) $ ({\bf Left}) and $\Delta \phi (ll, \met)$ ({\bf Right})  distributions. Only basic cuts have been imposed on these distributions. Even though these two plots are highly correlated, $\Delta \phi (ll, \met)$ exhibits a better signal - background separation.  That is because the pseudo-rapidity information of the \MET vector is not available at the LHC. Due to the similar behavior of the reducible backgrounds with their corresponding irreducible background, their contribution is ignored in this plots. }
 \label{fig:Deltaphi}
\end{figure}

 Another variable that is useful in signal-background discrimination is the azimuthal angle between the dilepton and the net missing energy vector $ \Delta \phi (\ell \ell, \met)$. Since leptons and neutrinos come from $X$, we expect to see some angular correlations between the leptons and the \MET vector. Because we do not know the pseudo-rapidity of the transverse missing energy (\MET) vector, the distribution of azimuthal angle provides a better discrimination than the total separation. A comparison between the two distributions of $ \Delta R (\ell \ell, \met)$ and $\Delta \phi (\ell \ell, \met)$ is presented in Fig~\ref{fig:Deltaphi}. For the benchmarks with $\mx \leq 50 \gev$,  the optimum cut seems to be $\Delta \phi (\ell \ell, \met) \lesssim 0.5$. 

\begin{figure}[t!]
 \centering
    \includegraphics[width=.5\textwidth]{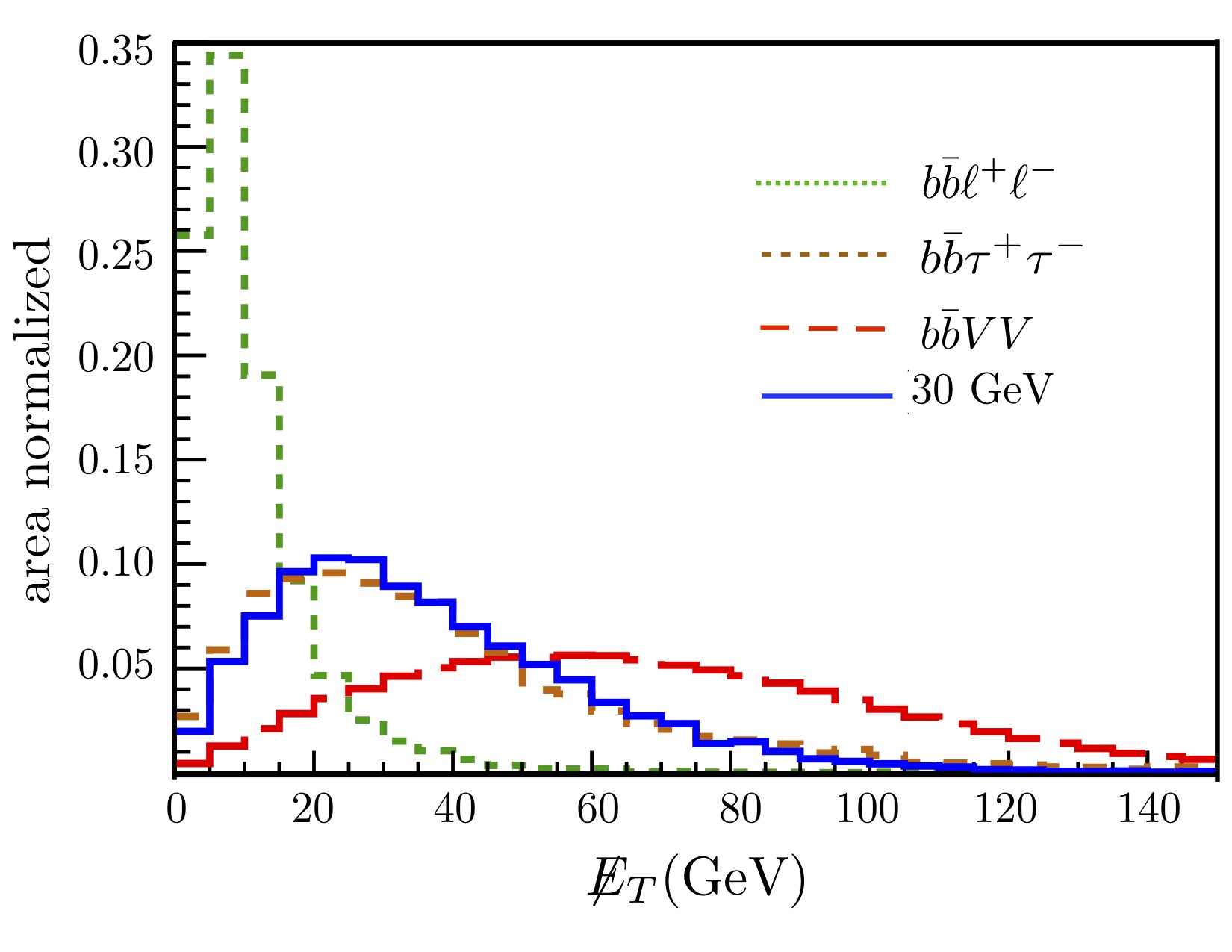}
  \caption{The area-normalized distribution of the transverse missing energy after the basic cuts is shown. Similar to previous distribution plots, we have ignored the reducible background's contribution. }
 \label{fig:met}
\end{figure}

Finally, we use the \MET distribution to further discriminate between the signal and the background. As shown in Fig.~\ref{fig:met}, the signal favors low \MET regime. That is because $b\bar b X$ production is maximum near threshold, where $X$ is almost stationary. The two neutrinos are, therefore, almost back-to-back, resulting in low \MET in the signal. Thereby, we can impose an upper limit on \MET, to suppress $\ttbar$ and $b \bar b \tau^+_\ell \tau^-_\ell$ backgrounds.  

Unfortunately, one background that favors low \MET is $\bl$, because its \MET is mostly a result of mismeasurement. To reduce this background, we must impose a lower bound on \MET in addition to the upper bound. The exact upper and lower \MET cuts were determined using our simulated events and adjusted to optimize the significance; the specific values for each of the benchmarks is presented in Appendix~\ref{app:cutflow}, however the cut on \MET is roughly $10 \lesssim \met \lesssim 70 \gev$.
 
\begin{table}[h!]
\begin{center}
\begin{tabular}{l c c c c | c  c   }
\hline
\hline
cuts &{$X_\mu$}  &   \multicolumn{3}{c}{SM backgrounds (fb)}&  \\
\hline
 &   $(30 \ \text{GeV}, 0.02) $&$\begin{matrix} b\bar b \ell^+ \ell^- \\  {c\bar c \ell^+ \ell^-} \\ {j_l j_l \ell^+ \ell^-}  \end{matrix} $& $\begin{matrix} b\bar bVV \\ {c\bar c VV} \\ {j_l j_l VV} \end{matrix}  $&  $ \begin{matrix} b\bar b \tau^+ \tau^- \\ {c\bar c \tau^+ \tau^-} \\  {j_l j_l \tau^+ \tau^-} \end{matrix}$& $\left.\text{significance}\right|_{\Lc = 100 \fb^{-1}}$   \\
 \hline
basic selection& 8.13 &$\left.\begin{matrix} 254.3 \\ 20.3 \\ {8} \end{matrix} \right.$&  $\left.\begin{matrix} 1362 \\ {2} \\ {1.8} \end{matrix} \right. $& $ \left.\begin{matrix} 27.8  \\ {2.4} \\ {2} \end{matrix} \right.$ & 0.005 \\
\hline
$ m_{\ell\ell} < 25 \ \text{GeV}$  &7.9  &$\left.\begin{matrix} 2.6 \\ {0.85} \\ {0.5} \end{matrix} \right.$&  $\left.\begin{matrix} 12.44 \\ {0.2} \\ {0.18} \end{matrix} \right. $& $ \left.\begin{matrix} 0.16\\ {0.06} \\ {0.02} \end{matrix} \right.$ & 5.77  \\
\hline
$\Delta \phi (\ell \ell, \met ) < 0.53 $  & 6.4  &$\left.\begin{matrix} 0.3 \\ {0.09} \\ {0.2} \end{matrix} \right.$&  $\left.\begin{matrix} 0.44 \\ {0.0} \\ {0.0} \end{matrix} \right. $& $ \left.\begin{matrix} 0.13 \\ {0.04} \\ {0.02} \end{matrix} \right.$ &40.0 \\
\hline
$14 < \met < 50 $  &5.48 &$\left.\begin{matrix} 0.09\\ {0.02} \\ {0.1} \end{matrix} \right.$&  $\left.\begin{matrix} 0.3 \\ {0.0} \\ {0.0} \end{matrix} \right. $& $ \left.\begin{matrix} 0.07 \\ {0.02} \\ {0.02} \end{matrix} \right.$ &49.0 \\
\hline
efficiencies & $67.4\%$  &$\left.\begin{matrix} 0.03\% \\ {0.01\%} \\ {0.1\%} \end{matrix} \right.$&  $\left.\begin{matrix} 0.02\% \\ {0.0\%} \\ {0.0\%} \end{matrix} \right. $& $ \left.\begin{matrix} 0.3\% \\ {0.8\%} \\ {1\%} \end{matrix} \right.$ &  \\
\hline
\hline
\end{tabular}
\caption{The effect of each cut on the signal benchmark with $\mx = 30 \gev$ and  each of the backgrounds.}
\label{tab:eff}
\end{center}
\end{table}
To quantify the sensitivity of our search, we follow the conventional definitions:
\begin{align}
 S \equiv \text{Luminosity} \times& \sigma( p p \to b \bar b \tau^+_\ell  \tau^-_\ell )_{\rm X}  \times \kappa_{\text{signal}}\nonumber \\
 B\equiv \text{Luminosity} \times& \left\{\sigma( p p \to b \bar b V_{\ell+ \tau_\ell }  V_{\ell+ \tau_\ell } )_{\rm SM}   \times \kappa_{\ttbar}\right.
\nonumber\\
&+ \sigma (p p \to b \bar b \tau^+_{_\ell} \tau^-_{_\ell} )_{SM}\times \kappa_{\btau}\nonumber\\
 & + \left.\sigma ( p p \to b \bar b \ell^+ \ell^-)\times \kappa_{\bl} \right\}.
 \label{eq:num} 
 \end{align}
In other words, $S$ and $B$ are respectively the number of signal and total background events (scaled to NLO rates) that we expect to observe at the LHC for a given luminosity and our cut flow. Using these, we quantify the discovery potential of our analysis using the significance, defined --following Ref.~\cite{1404.0682} -- as:
 \beq
 \text{Significance} = \frac{S}{\delta B} = \frac{S} {\sqrt{B + \sum_i \lambda_i^2 B^2 }},
 \label{eq:sig}
 \eeq
where $\lambda_i$ represent the systematic uncertainties associated with each background. We used the following values for $\lambda_i$, taken from the CMS leptonic $\ttbar$ search~\cite{1603.02303, Aaboud:2016pbd}:  $\lambda_{\ttbar} = 5 \%$, $\lambda_{\btau} = 10 \%$, and $\lambda_{\bl} = 15 \%$. Finally, it is important to note that we have ignored the effect of pile-up in our analysis. Including the effect of pile-up will likely affect the lower bound on \MET. Specifically, it will effect the contribution of the $\bl$ process in the background. However, the cross section of $\bl$ even before imposing the \MET cut is already much smaller than other processes, and thus we do not expect that a small change in its cross section to alter our results significantly.

\begin{figure}[t!]
 \centering
  \includegraphics[width=.75\textwidth]{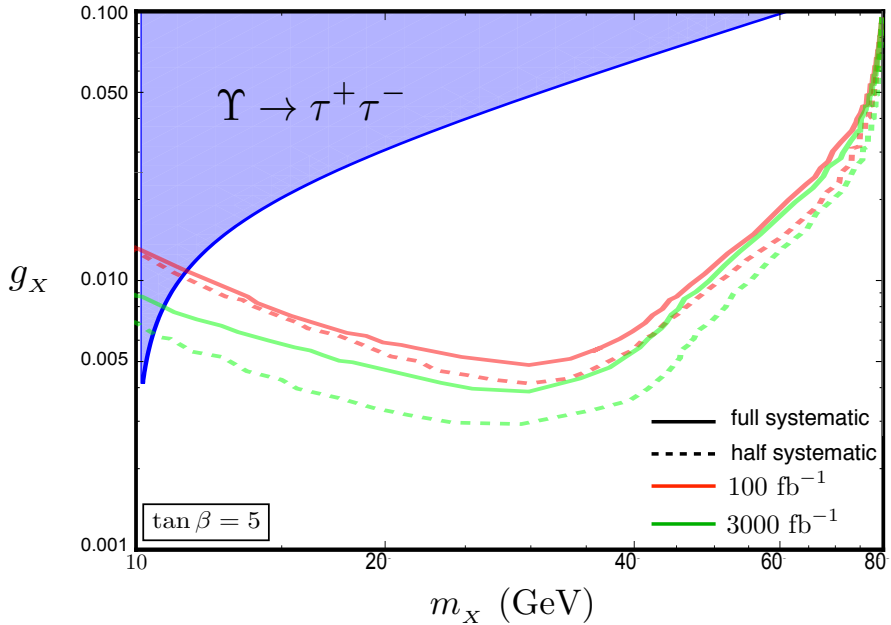}
  \caption{By imposing an optimized cut on the mentioned variables, the LHC can probe the indicated parameter space up to $3 \ \sigma$ significance, with $100 \fb^{-1}$ integrated luminosity. The solid red line represents the significance with the systematic uncertainties mentioned in the section. The dashed red line is the significance when the systematic uncertainties are half. The green lines show the projected sensitivity at HL-LHC with $3\, \ab^{-1}$ integrated luminosity. The bound from $ \Upsilon \to \tau^+ \tau^-$ at BaBar~\cite{1002.4358} in shown in blue. Due to the low efficiency of the trigger cut, our sensitivity to very light $m_{X}$ is weak; for $20  \gev<  m_{_X} < 40 \gev$ we have our maximum sensitivity, then the sensitivity drops again as $m_{X}$ increases due to a combination of a lower signal cross section and a decline in the discriminatory power of the cuts. In Sec.~\ref{chisq}, we will motivate and study some variables that enhance the sensitivity to heavier $m_{X}$ ($> 50\, \gev$). The presented bounds are for $\tan \beta = 5$. However, for any $\tan \beta >1 $, the sensitivity is almost the same.} 
 \label{fig:constraint}
\end{figure}

For each $\mx$ benchmark, the cuts on the $m_{\ell\ell}, \Delta\phi(\ell\ell, \slashed E_T)$ and $\slashed E_T$ distributions have been optimized to yield the largest significance (Eq.~\eqref{eq:sig}). Once the cuts have been optimized, we use Eq.~\eqref{eq:gx} to extrapolate the analysis to other $\gx$ values and trace out contours of a desired significance, as shown in Fig.~\ref{fig:constraint}. The red lines are the bounds with (roughly) the current luminosity of the LHC -- $100 \fb^{-1}$ -- while the green lines are the projected sensitivity with the full luminosity of HL-LHC ($3\,\ab^{-1}$). The solid lines present the  3 $\sigma$ (significance as defined in Eq.~\eqref{eq:sig} = 3) exclusion bound assuming the full systematical uncertainties mentioned earlier, and the dashed lines are $3\ \sigma$ significance when the systematic uncertainties are cut into half.

As we can see from Fig.~\ref{fig:constraint}, the LHC can probe a region of the parameter space that is out of reach of other current experiments. The LHC bounds are best in the mass range $ 20 - 40 \gev$. The constraints on lighter $X$ are milder, due to the low dilepton trigger efficiency and a relatively lower b-tagging efficiency, and the sensitivity for heavier $X$ drops because the distribution of the signal and backgrounds become more similar, and so the cuts become less efficient. 

Above $50\, \gev$, the limits worsen quickly. Therefore, it is worth exploring if there are any additional variables that can improve the bounds in the mass range $ 50 \gev < \mx < \mz$. The most troublesome background in this mass range is $\ttbar$, and the most important contribution to $\ttbar $ comes from $t\bar t$. The goal of the next sub-section is to investigate some kinematic handles that specifically target the $t \bar t$ background.

\subsubsection{ Further Separation of the Signal from the $\ttbar$ background, for $\mx \sim \mz$ }
\label{chisq}
\begin{figure}[t!]
 \centering
\includegraphics[width=.5\textwidth]{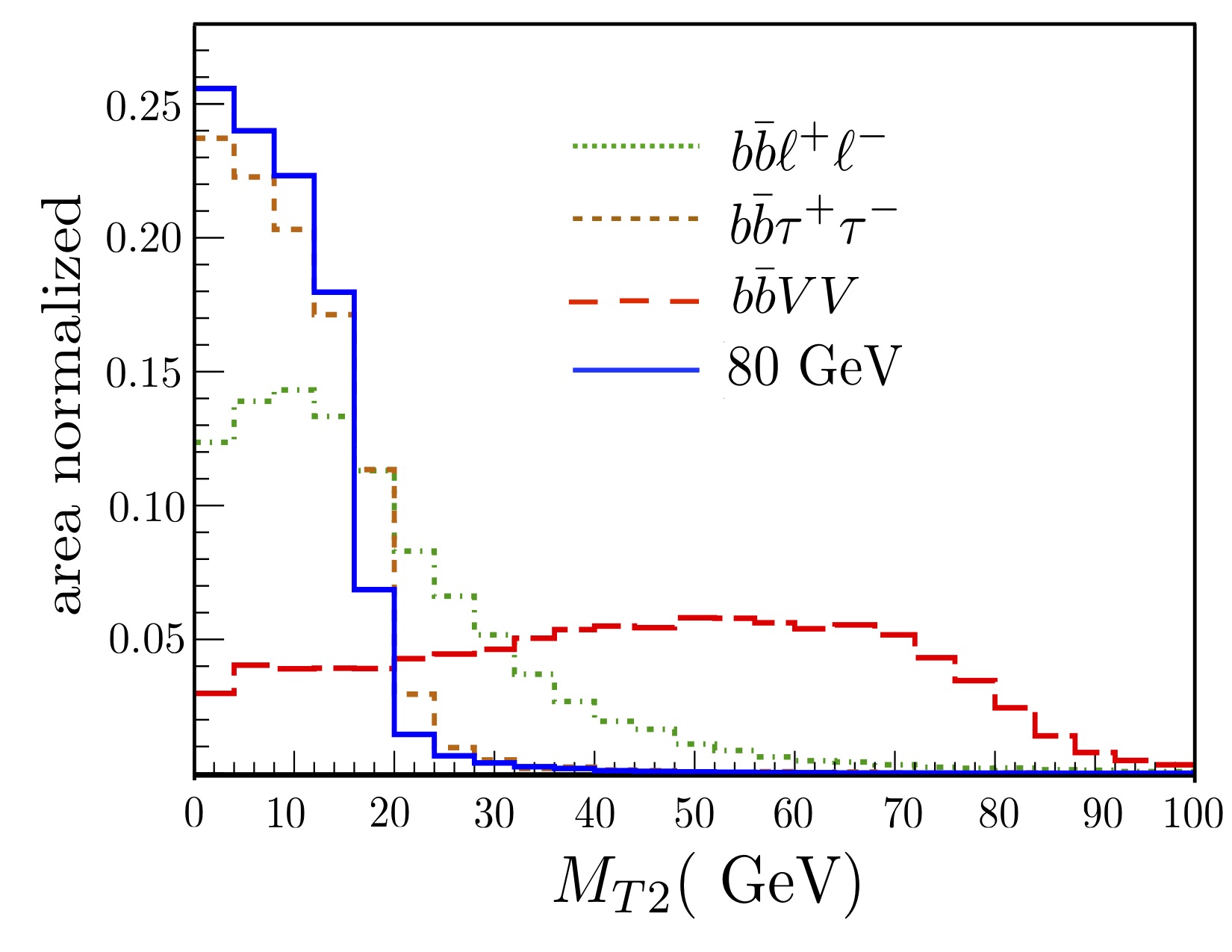}~
 \includegraphics[width=.51\textwidth]{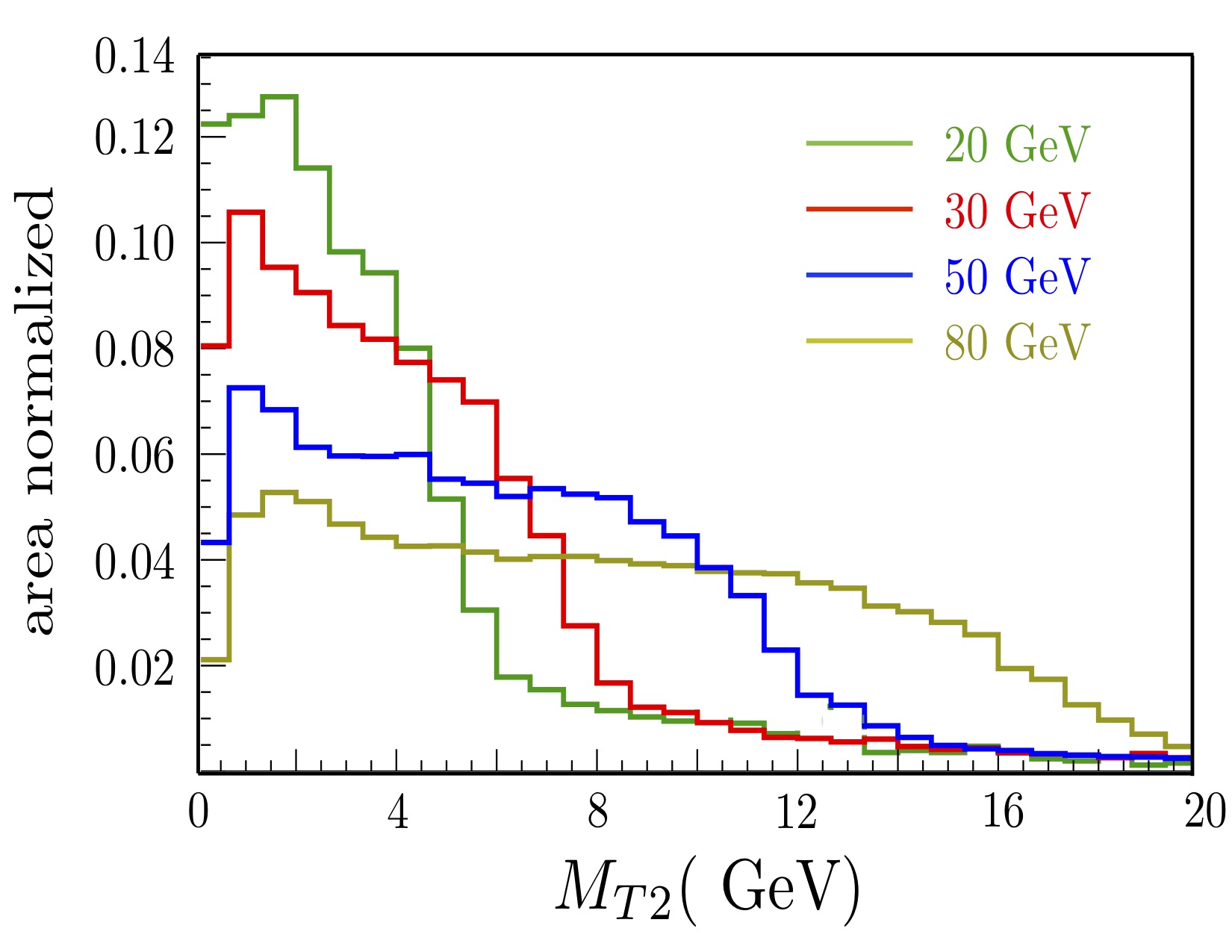}
  \caption{The area-normalized distribution of various benchmarks {\bf(left)} and the signal-backgrounds comparison {\bf (right)} with respect to $M_{T2}$ is shown. The signal prefers small value of $M_{T2}$, whereas the $\ttbar$ backgrounds has a broad feature-less distribution.} 
 \label{fig:mt2}
\end{figure}

The dominant background to our signal is $ t\bar t \to b \bar b W^+_{\ell, \tau_{\ell}}W^-_{\ell, \tau_{\ell}}$. This background has some specific features that can help us separate this process from the signal. For example, the leptons and neutrinos in $t \bar t$ come from $W$ decays, whereas in the signal, they are due to $\tau$ decays. Therefore, $M_{T2}$, defined as 
\beq
M_{T2} = \text{min}_{\nu^1_T + \nu^2_T = \met} (\text{max} (M_T(\ell^+, \nu_1),M_T(\ell^-, \nu_2))),
\eeq
with $\nu^i_T$ being the transverse momentum of the either sources of missing energy, should show a decent separation between the signal and the $t\bar t$ background. Since we are particularly interested in enhancing the sensitivity for $\mx \sim \mz$, in the left panel of Fig.~\ref{fig:mt2}, we compare the distribution of the benchmark with $\mx = 80\, \gev$ with the backgrounds. As expected, $M_{T2}$ of the signal prefers small values ($\lesssim 20 \gev$), while $M_{T2}$ for the $t\bar t$ background $\sim m_W$. The distribution of $M_{T2}$ for various benchmarks, after basic cuts only (so no $m_{\ell\ell}, \Delta \phi(\ell\ell,\slashed E_T)$, etc.), is also shown in the right panel of Fig.~\ref{fig:mt2}. Even though the separation of the signal from the $\ttbar$ background is more visible for lighter $X$, $M_{T2}$ does not improve the bounds for $\mx < 50 \gev$ compared to the combination of kinematic cuts introduced in the previous section.
For heavy $X$, however, the efficiency of a cut on $\Delta \phi (\ell \ell, \met)$ and $\met$ distributions is not as efficient as a cut on $M_{T2}$. By requiring $M_{T2} < 16 \gev$, the significance goes up by a factor of 3 for $\mx = 80 \, \gev$, and the mild discrimination in the $\Delta \phi (\ell \ell, \met)$ and $\met$ distributions fade off. Therefore, we can no longer impose an efficient cut on  $\Delta \phi (\ell \ell, \met)$ and $\met$ distributions. 

\begin{figure}[t!]
 \centering
 \includegraphics[width=.6\textwidth]{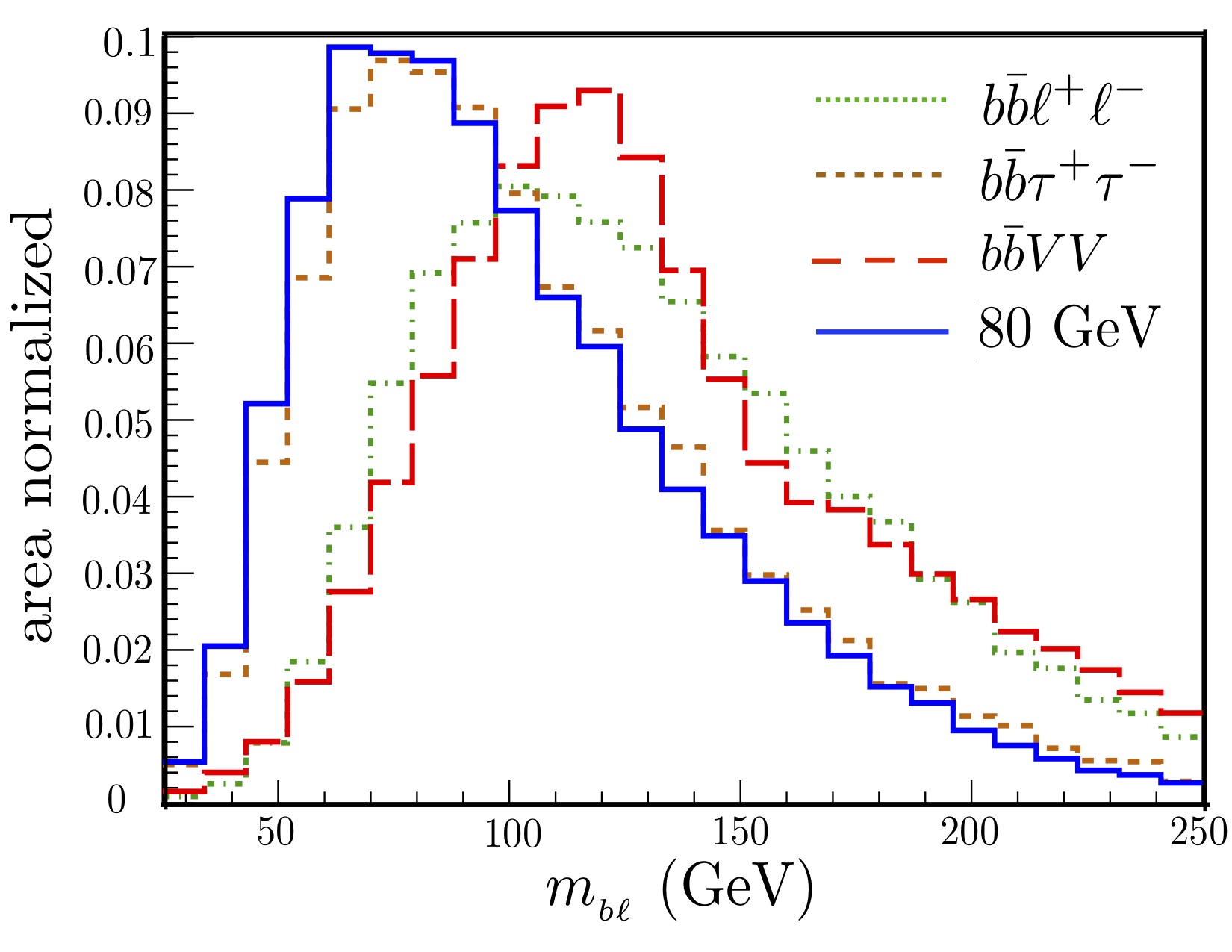}
  \caption{The invariant mass of $b$ and lepton that minimizes $(m_{\ell_i b_k} -m_t)^2 + $ $(m_{\ell_j b_n} -m_t)^2$ is shown here. The distribution of both reconstructed tops are very similar, and so we are showing only one combination. The signal distribution, shown in solid blue, is for the benchmark $\mx = 80 \gev$. The most important background is $t \bar t$, shows is shown in dashed red.  } 
 \label{fig:bl}
\end{figure}

Another attribute of the $t\bar t$ background is that there is an intimate relationship between $b$ jets and leptons: $m_{bl} \lesssim m_t$. In the signal, on the other hand, such correlation does not exist, and $m_{bl}$ can take any arbitrary values. To take advantage of this difference, we study $m_{b\ell}$ in Fig.~\ref{fig:bl}, where the $b- \ell$  is one of the combinations that minimizes $(m_{\ell_i b_k} -m_t)^2 + (m_{\ell_j b_n} -m_t)^2)$, with $ i \neq j$ and $k \neq n$. The distributions for $m_{b\ell}$ for the backgrounds and a $\mx = 80\,\gev$ benchmark are shown below in Fig.~\ref{fig:bl}. We can see that there is a modest separation\footnote{According to the distributions in Fig.~\ref{fig:bl}, the signal mostly resides in $m_{b\ell} \lesssim 200 \gev$, and thus the CMS search for third generation leptoquarks~\cite{CMS:2018pab,Khachatryan:2014ura} with $m_{b\ell} > 250 \gev$ does not have a noticeable sensitivity to our signal.} between the $t \bar t$ background and other processes. A cut on this distribution can enhance the significance by $15\%$ for the benchmark $(\mx, \gx) = (80 \gev, 0.02)$.

\begin{figure}[t!]
 \centering
  \includegraphics[width=.5\textwidth]{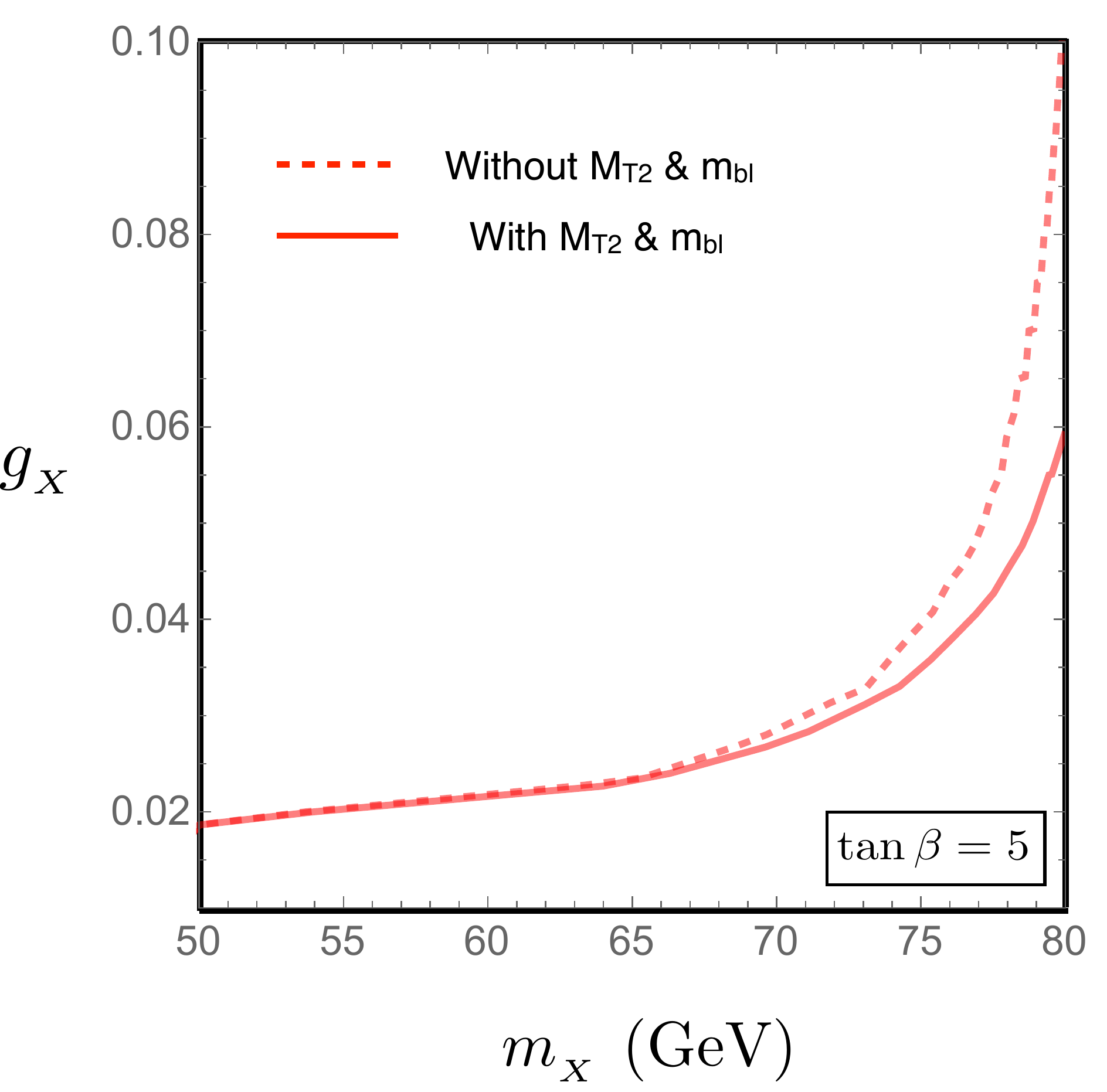}
  \caption{The contribution of $M_{T2} $ and $m_{bl}$ to improving the sensitivity for $\mx = 80 \gev$. The dashed line indicates the 3 $\sigma$ significance with only the basic cuts and $25 < m_{\ell \ell}< 60 \gev$ with $100 \fb^{-1}$ integrated luminosity. The solid line is the 3 $\sigma$ significance with the same amount of luminosity with the basic cuts + $25 < m_{\ell \ell} < 60 \gev$ + $ m_{T2} < 16 \gev$ + $m_{b\ell} < 100 \gev$. The extra cuts improve the sensitivity by roughly factor of $1.7$.
  } 
 \label{fig:MT2cont}
\end{figure}

To improve on this guess, we tried finding the neutrino momenta by reconstructing the $W$ and top mass. In particular, we scanned through all possible values of momenta that give the smallest $\chi^2$, defined as:
\beq
\chi^2 =  \frac{ (m_{\ell_i \nu_1}^2 - m_{_W}^2)^2 }{\sigma_{_W}^4} +  \frac{ (m_{\ell_j \nu_2}^2 - m_{_W}^2)^2 }{\sigma_{_W}^4} +  \frac{ (m_{b_k \ell_i \nu_1}^2 - m_{t}^2)^2 }{\sigma_{t}^4} +  \frac{ (m_{b_n \ell_j \nu_1}^2 - m_{t}^2)^2 }{\sigma_t^4},
\eeq
where $\sigma_{W}$ and $\sigma_t$ are arbitrary values we can use to enhance our discrimination. However, regardless of the values of $\sigma_{t,W}$, this method did not improve the signal-background discrimination. Therefore, the only cuts that could improve our sensitivity to $\mx \sim \mz$ were $M_{T2} < 16 \gev$ and $m_{bl} < 100 \gev$. With these cuts, more than $94\%$ of the background is removed, while almost $50 \%$ of the signal is preserved. The effect of these cuts on the significance is shown in Fig.~\ref{fig:MT2cont}. Since the cross section is proportional to two factors of the coupling ($\sigma \propto \gx^2$), improving the limit by a factor of 4 translates into an improvement in the coupling by a factor of 2.

Having exhausted the cut-based search strategies for light $\mx$, we now turn to  $\mx > \mz$.
 In this regime, the $\tau$s coming from the decay of $X$ gauge boson are expected to be energetic enough such that the resulting lepton from one of the $\tau$s can pass the single lepton trigger with high efficiency. Therefore, we study the semi-leptonic $b \bar b \tau_h \tau_\ell$.

\subsection{ Heavy $X$: $\mx >  \mz$ }
\label{sec:heavy}

The cross section for $X$ production falls as $\mx$ increases. To compensate for the lower cross section, for $\mx > \mz$ we shift final states to semi-leptonic $\tau$s (one tau decays to leptons, one to hadrons) to take advantage of the higher branching ratio of $\tau$ to hadrons.
The SM backgrounds we need to be concerned about are 
\begin{eqnarray}
1.)\quad && pp \rightarrow\, b \bar b\  W_{\tau_{_h}} W^\pm_{\ell+\tau_{_\ell}}  \nonumber\\
2.)\quad && pp\rightarrow\,  b \bar b\  \tau_h \tau^\pm_\ell  \nonumber\\
3.)\quad &&  pp \rightarrow\, b \bar b  W^\pm_{\ell+\tau_\ell} + \text{jets} \nonumber\\
4.)\quad && pp\rightarrow\, b \bar b \ (Z/\gamma)_{\ell+ \tau_\ell} + \text{jets} , 
\label{eqn:backgrounds2}
\end{eqnarray}
where, $\ell = e, \mu$, and $\tau_\ell$ refers to the leptonic decay of a $\tau$. Similarly, $h$ represents the hadronic decay. Only the first two backgrounds mentioned in Eq.~\ref{eqn:backgrounds2} are irreducible. As hadronic taus can be faked by `normal' jets ($= j_l+ c+\bar c+ b + \bar b$), we must include lepton + jet backgrounds such as 3.) and 4.) above\footnote{Backgrounds 3.) and 4.) in Eq.~\eqref{eqn:backgrounds2} are separated as they contain different numbers of charged leptons. The third background has one lepton, and the fourth background contains two leptons, with some probability that one of the leptons fall outside of the acceptance and manifests itself as missing energy. }. To estimate the backgrounds with fake taus, we rely on the built-in tau identification algorithm in Delphes, where we input matched samples\footnote{We use MLM matching with Madgraph5 + PYTHIA8, with ${\tt xqcut} = 20$~\cite{Alwall:2007fs}. We have included up to $2$ jets, e.g $pp \rightarrow\, b \bar b  W^\pm_{\ell+\tau_\ell} + 0,1,2\, \text{jets}$. The matched/merged cross sections are then rescaled to the $+0\,\text{jet}$ NLO values.}.

As in the previous section, we must also consider backgrounds where a charm jet or a light quarks/gluon jet ($j_l$) is mis-identified as a b-jet. For example:
\beq
3.) pp \to b \bar b\, \ W^\pm_{\ell+\tau_\ell} + \text{jets}  \Rightarrow \left\{ \begin{array}{c} pp \to c\bar c\ W^\pm_{\ell+\tau_\ell} + \text{jets}  \\ pp \to j_l\,j_l W^\pm_{\ell+\tau_\ell} + \text{jets}  \end{array}\right. ,
\eeq
where `$+\, \text{jets}$' is treated as up to $+0-2$ jets.\\ 

To capture the interesting events, we impose the following conditions: 
\begin{itemize}
\item[] -- Each event must include exactly one charged lepton that passes the single lepton trigger:  $p_T(\ell_1) > 27 \ (24)\, \gev$ if the lepton is electron (muon).  We also require $|\eta (\ell)|  < 2.5$. 
\item[] -- We require one $b$-jet possessing $p_T (j) > 50\, \gev$, and $|\eta(j)| < 2.5$. As in the previous section, we  use the {\tt Delphes}~\cite{deFavereau:2013fsa} b-identification algorithm to tag a $b$ jet. 
\item[] --  Every event must contain one tau-tagged hadronic jet, $p_T (j) > 50\, \gev$, and $|\eta(j)| < 2.5$. As with $b$-jets, we rely on the built-in algorithm in {\tt Delphes}~\cite{deFavereau:2013fsa}. We find the $\tau$ tagging efficiency is roughly $40\%$ for correctly identifying a hadronic $\tau$ with an $\sim 0.3\%$ risk of misidentifying a normal jet as a hadronic $\tau$, for the processes being considered here. 
\item[] -- In addition to the $b$-jet and $\tau$ jet, the event may contain at most one additional jet, $p_T (j) > 50\, \gev$, and $|\eta(j)| < 2.5$. The separation between each jet, as well as  the separation between all jets and the lepton must be greater than $0.4$. 

\end{itemize}

 Due to the presence of multiple jets in our final state of interest, one might expect the main backgrounds come from tau/b-misidentified jets. However, after requiring the basic cuts mentioned, 
the largest background is the irreducible $  b \bar b\  W_{\tau_{_h}} W^\pm_{\ell+\tau_{_\ell}}$, 120 fb at NLO. The other sizable backgrounds are $ b \bar b\  \tau_h \tau^\pm_\ell$,  (3.5 fb), $b \bar b  W^\pm_{\ell+\tau_\ell} + \text{jets}$ (3 fb), and $ b \bar b \ (Z/\gamma)_{\ell+ \tau_\ell}   + \text{jets}$ (0.3 fb). All other backgrounds are negligible, $ \ll 0.1 \fb$.

To enhance the sensitivity of the signal further, we studied various kinematic distributions including $M_{T2}$, $m_{\ell j_i}$ -- where $j_i$ is any of the jets in the final state, the separation between the lepton and the jets $\Delta R (\ell, j_i)$, the difference in the azimuthal angle between any two visible objects in the final state, as well as the $p_T$ of each of the visible final states. Some of these distributions show a small difference between the signal and background, but none of them have a considerable effect on their own. Therefore, for this initial study, we will ignore the impact of these other distributions and quantify the sensitivity using the basic cuts alone. A multivariate analysis may be able to harness the slight differences across several kinematic distributions and yield increased sensitivity. Such an approach would be interesting to pursue, but is beyond the scope of the current work. However, as the difference in the distributions are very small, we expect the sensitivity gains achieved by a MVA to be $\mathcal O(1)$ in the cross section and not orders of magnitude.

\begin{figure}[t!]
 \centering
  \includegraphics[width=.7\textwidth]{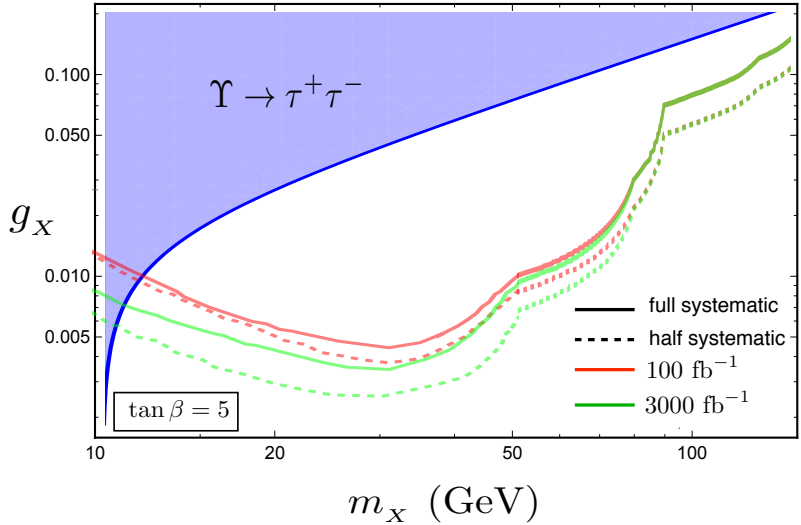}
  \caption{he reach up to $3 \ \sigma$ significance after $100 \fb^{-1}$ integrated luminosity is shown. The current bound coming from $ \Upsilon \to \tau^+ \tau^-$ at BaBar~\cite{1002.4358} excludes the blue region.} 
 \label{fig:constraint2}
\end{figure}

 Using a similar definition of the signal and background as in Section~\ref{sec:light}:
\begin{align*}
 S &\equiv \text{Luminosity} \times \sigma( p p \to b \bar b \tau^+ \tau^- \to  b \bar b \ell^\pm  j  +2 \nu  )_{X}  \\
 B& \equiv \text{Luminosity} \times \left(\sigma( \bar b\  W_{\tau_h} W^\pm_{\ell+\tau_\ell} )_{\rm SM} + \sigma (p p \to b \bar b \tau^\pm_h \tau^\mp_{\ell})_{SM} + \sigma ( p p \to b \bar b (W^\pm /Z/\gamma)_{\ell} + \text{jets} \right), 
 \end{align*}
 with 
 \begin{align*}
 \text{Significance} &= \frac{S}{\delta B} = \frac{S} {\sqrt{B +  \lambda^2 B^2}}.
 \end{align*}
 and extrapolating to all values of $g_X, \tan{\beta}$ using Eq.~\eqref{eq:gx}, we can chart significance contours.
 In the $\mx > \mz$ region, the main background is the irreducible background $ p p \to t \bar t \to  b \bar b\  W_{\tau_{_h}} W^\pm_{\ell+\tau_{_\ell}}$, which has a systematic uncertainty of $\lambda = 12\%$~\cite{Aaboud:2018mjh}. We will assume the same uncertainty on the rest of the backgrounds as well, though an $O(1)$ change in the systematic uncertainties of other backgrounds does not affect the results significantly.  Assuming $ 100 \fb^{-1}$ integrated luminosity, we can exclude up to $\gx \gtrsim 0.07$ for $\mx = 100 \gev$, and $\gx \gtrsim 0.2$ for $\mx = 150 \gev$ up to $3 \sigma$ significance, illustrated in Fig.~\ref{fig:constraint2}. This is about a factor of $2-4$ improvement over previous constraints. Even though the total background of the semi-leptonic $b \bar b \tau_h \tau_\ell$ after the basic cuts is much smaller than that of the fully leptonic $b \bar b \tau_\ell \tau_\ell$, the constraints in the $\mx \ll \mz$ region are much stronger. That is because in the $ \mx \ll \mz$ region, the kinematic distributions of the signal have sharp features that distinguishes it from the background. For larger masses, however, the distributions broaden and lose their sharp features and thus separating the signal from the background is more challenging.

In general, a dedicated search at the LHC can improve the bounds by a factor of $2-10$. These results can be achieved by studying simple kinematic distributions. With the usage of a more advanced technique like an MVA, we might obtain even better results. Moreover, we have stopped our search at $\mx < 2 m_W$. The bounds on a larger $\mx$ will depend on some parameters in the scalar potential that we have ignored for our study (e.g, mixing between the scalars). If $\mx > 2 m_t$, the decay of $X$ to a pair of top quarks enjoys a significant probability as well as small  background due to the large number of final state particles. These searches have already received some attention in several phenomenological studies~\cite{Hill:1991at,Hill:1993hs,Hill:1994hp,Harris:2011ez,Rosner:1996eb,Lynch:2000md,Carena:2004xs,Choudhury:2007ux,Khachatryan:2015sma,CMS:2018ohu,Aaboud:2018mjh,Sirunyan:2017uhk,Sirunyan:2017yar,Cerrito:2016qig,Arina:2016cqj,Pedersen:2015knf,Fox:2018ldq}. The constraints obtained by these studies can be recasted according to our choice of model parameter values.  

\section{ Conclusion }
\label{conclusion}

In this work, we explored the LHC potential to constrain $X$, the gauge boson associated with a spontaneously broken $\U$ symmetry. $\U$ symmetry is one of the simplest extensions of the SM, which was first proposed to explain the flavor alignment of the third generation of quarks.  While $X$ only interacts with the third generation of fermions in the interaction basis (at tree-level), flavor non-universal couplings are generated once we rotate to the mass basis. These flavor-violating effects can be mitigated with certain charge assignments and coupling assumptions, but strong constrains from low-energy experiments persist for $\mx \le 5\,\gev$.  

To hunt for heavier $X$, which are free from low-energy bounds, we developed two dedicated LHC search strategies based on $pp \to X, X \to \tau^+\tau^-$, a production and decay path that yields a high rate and numerous kinematic handles to suppress SM backgrounds. Following Ref.~\cite{1705.01822}, we assume all scalars related to $\U$ breaking and the right handed neutrino are heavy, and focus on $\mx \le 2\,m_W$ since this decouples the $X$ phenomenology from any mixing in the scalar sector.

For $\mx < m_Z$, we find the optimal channel is where both taus decay leptonically. Using a combination of simple kinematic variables, such as $m_{\ell \ell}, \Delta \phi (ll, \met)$ and \MET, we find that couplings as low as $\gx > 0.005$ for $ \mx \sim 20 \gev$ could be probed at $3 \ \sigma$ sensitivity given $100 \fb^{-1}$ integrated luminosity (roughly the current total LHC-13 luminosity). For heavier masses, the bounds are not as strong: $\gx > 0.05$ for $\mx \sim 80 \gev$ probed at $3\,\sigma$ with the same amount of data. Extrapolating these bounds to the full HL-LHC luminosity of $3\, \text{ab}^{-1}$, we expect a further increase by a factor of 2 in the sensitivity (or $\sqrt 2$ in $\gx$).

For $\mx > m_Z$, we find the semi-leptonic tau channel ($\tau^+_h\tau^-_\ell$ + $\tau^+_\ell\tau^-_h$) outperforms the fully leptonic mode, however the number of pronounced kinematic differences between the signal and the dominant background ($pp \to t\bar t$) shrinks substantially. For $100 \fb^{-1}$, we find the $3\,  \sigma$ exclusion limit reaches  $\gx \gtrsim 0.1$ for $\mx = 100 \gev$, and $\gx \gtrsim 0.2$ for $\mx = 150 \gev$. Both the low-mass and high-mass search strategies relied on cut-and-count methods and it would be interesting to explore what improvements multivariate techniques can squeeze out.

\section*{Acknowledgments}
\label{sec:ack}

We are particularly thankful to S. Chenarani for the numerous insightful conversations. We would like to also thank H. Hesari, S. Khatibi, J. H. Kim, M. Paktinat, and F. Rezai for their useful comments.  We are grateful to CERN theory group for their hospitality. The work of AM was partially supported by the National Science Foundation under Grants No. PHY-1820860.

\appendix

\section{The cut flow of benchmarks with $\mx < \mx$}
\label{app:cutflow}
 The cut flow on each of the benchmarks is shown here. The quoted cross sections are at NLO, even though the events are LO. We generated $10^6$ events for the signal, $\bl$, and $b \bar b \tau_\ell^+ \tau_\ell^-$ processes. Due to the higher cross section of $\ttbar$, we generated $ 2 \times 10^6$ events for this process. 
 In all of the benchmarks studied here, $m_{\ell \ell}$ proved to be a useful variable in distinguishing the signal. For $\mx \leq 50 \gev$, we used $\Delta \phi (\ell \ell, \met)$ and $\met$ to further distinguish the signal, and for $\mx = 80 \gev$, we found $M_{T2}$ to be a more useful variable. Each cut has been optimized such that it gives the highest significance, defined in Eq.~\eqref{eq:sig}. 
\begin{table}[H]
\begin{center}
\begin{tabular}{l c c c c c c c c }
\hline
\hline
cuts &{$X_\mu$}  &   \multicolumn{3}{c}{SM backgrounds (fb)}&$\left.\text{significance}\right|_{\Lc = 100 \fb^{-1}}$  \\
\hline
 &   $(10 \ \text{GeV}, 0.02) $&$b\bar b \ell^+ \ell^- $& $\ttbar$&  $b\bar b \tau^+ \tau^-$  &  \\
\hline
 basic selection  & 0.61 & 282.6 & 1365.8 &32.2 & 0.005    \\
$ m_{\ell\ell} < 10 \ \text{GeV}$  &0.5& 0.5& 1.55& 0.03 & 2.3   \\
$\Delta \phi (\ell \ell, \met ) < 0.5 $  & 0.3 & 0.05& 0.26& 0.03 &4.6 \\
$ 5 \gev < \met < 70 \gev $  & 0.3 & 0.03& 0.21& 0.03 &5.3\\
efficiencies & $50\%$ & $0.01\%$ & $0.016\%$& $0.1\%$ &     \\
\hline
\hline
\end{tabular}
\caption{The cut-flow for $\mx = 10 \gev$. The cross section of this benchmark is small after the basic selection. However, the $m_{\ell \ell}$ cut efficiently enhances the signal to background ratio.   }
\end{center}
\end{table}

\begin{table}[H]
\begin{center}
\begin{tabular}{l c c c c c c }
\hline
\hline
cuts &{$X_\mu$}  &   \multicolumn{3}{c}{SM backgrounds (fb)}&$\left.\text{significance}\right|_{\Lc = 100 \fb^{-1}}$  \\
\hline
 &   $(20 \ \text{GeV}, 0.02) $&$b\bar b \ell^+ \ell^- $& $\ttbar$&  $b\bar b \tau^+ \tau^-$&   \\
\hline
  basic selection  & 2.27 &282.6 & 1365.8 &32.2& 0.02    \\
$ m_{\ell\ell} < 15 \ \text{GeV}$  &2.04 & 1.88 & 12.33 & 0.16 & 2.00 \\
$\Delta \phi (\ell \ell, \met ) < 0.5 $  & 1.69 & 0.23& 0.44& 0.13& 12.9 \\
$10 < \met < 70 $  & 1.67& 0.06& 0.30& 0.07 &21.0\\
efficiencies & $73.2\%$ & $0.02\%$ & $0.02\%$& $0.27\%$ &    \\
\hline
\hline
\end{tabular}
\caption{The effect of each cut on the signal with $\mx = 20 \gev$ and the backgrounds.  The $m_{\ell \ell}$ cut efficiently enhances the signal to background ratio. }
\end{center}
\end{table}

\begin{table}[H]
\begin{center}
\begin{tabular}{l c c c c c c }
\hline
\hline
cuts &{$X_\mu$}  &   \multicolumn{3}{c}{SM backgrounds (fb)}&$\left.\text{significance}\right|_{\Lc = 100 \fb^{-1}}$  \\
\hline
 &   $(30 \ \text{GeV}, 0.02) $&$b\bar b \ell^+ \ell^- $& $\ttbar$&  $b\bar b \tau^+ \tau^-$  &  \\
\hline
 basic selection  & 8.13 &282.6 & 1365.8 &32.2& 0.07    \\
$ m_{\ell\ell} < 25 \ \text{GeV}$  &7.9& 3.95& 12.82 & 0.24 & 5.77 \\
$\Delta \phi (\ell \ell, \met ) < 0.53 $  & 6.4 & 0.32& 0.44& 0.19& 40.0 \\
$13 < \met < 60 $  & 5.48 & 0.21& 0.31& 0.11&49.0 \\
efficiencies & $67.4\%$ & $0.04\%$ & $0.01\%$ & $0.3\%$  & \\
\hline
\hline
\end{tabular}
\caption{The effect of each cut on the signal with $\mx = 30 \gev$ and the backgrounds.  The $m_{\ell \ell}$ cut efficiently enhances the signal to background ratio.}
\end{center}
\end{table}


\begin{table}[H]
\begin{center}
\begin{tabular}{l c c c c c c }
\hline
\hline
cuts &{$X_\mu$}  &   \multicolumn{3}{c}{SM backgrounds (fb)}&$\left.\text{significance}\right|_{\Lc = 100 \fb^{-1}}$  \\
\hline
 &   $(50 \ \text{GeV}, 0.02) $&$b\bar b \ell^+ \ell^- $& $\ttbar$&  $b\bar b \tau^+ \tau^-$  &  \\
\hline
 basic selection  & 3.42 &282.6 & 1365.8 &32.2& 0.03 \\
$ 10 < m_{\ell\ell} < 35 \ \text{GeV}$  &2.88 &6.35 &163 & 5.58 & 0.27 \\
$\Delta \phi (\ell \ell, \met ) < 0.5 $  & 2.08&0.74& 7.07&3.38& 1.37 \\
$22 < \met < 75 $  &1.75& 0.03& 6.5 &0.3 & 3.5\\
efficiencies & $51\%$ & $0.01\%$ & $0.48\%$ & $1.17\%$  & \\
\hline
\hline
\end{tabular}
\caption{The effect of each cut on the signal with $\mx = 50 \gev$ and the backgrounds.}
\end{center}
\end{table}


\begin{table}[H]
\begin{center}
\begin{tabular}{l c c c c c c }
\hline
\hline
cuts &{$X_\mu$}  &   \multicolumn{3}{c}{SM backgrounds (fb)}&$\left.\text{significance}\right|_{\Lc = 100 \fb^{-1}}$  \\
\hline
 &   $(80 \ \text{GeV}, 0.02) $&$b\bar b \ell^+ \ell^- $& $\ttbar$&  $b\bar b \tau^+ \tau^-$  &  \\
\hline
 basic selection  & 2.8 & 282.6 & 1365.8 &32.2&  0.02   \\
$ 25 < m_{\ell\ell} < 60 \ \text{GeV}$  &2.51& 2.58 & 317.2 &21.25 &0.11\\
$M_{T2} < 16 \gev$ & 2.24 &  1.6  &41.3 &17.3& 0.30\\
$m_{b\ell} < 100 \gev$& 1.15 & 0.56 & 10.3 & 8.8 & 0.35\\
efficiencies & $41.1\%$ & $0.2\%$ & $0.8\%$ & $27.3\%$  & \\
\hline
\hline
\end{tabular}
\caption{The effect of each cut on the signal with $\mx = 80 \gev$ and the backgrounds. To further enhance the signal to background ratio, $M_{T2}$ and $m_{b\ell}$ is used. }
\end{center}
\end{table}



\section{Reproducing the CKM Matrix and Flavor Changing Interactions of $X$}
\label{app:interaction}

This model was first suggested in Ref.~\cite{1705.01822}, and the details of the model are somewhat complicated and lengthy. Rather than discussing all of the moving parts, we will focus on the generation of the CKM matrix and potential FCNC.

Because the third generation is charged under $\U$ while the first and second generations are not,  mixing among generations requires $\phi$. The full Yukawa interaction, including interactions with Higgs or $\phi$ and working in a basis with diagonal kinetic terms, can be written as:

\beq
\label{eq:yuk}
\L^q_{yuk} = \overline{\mathbf{Q}}_L
  \left(\begin{array}{ccc}
    y_{11}^u \widetilde H & y_{12}^u \widetilde H & y_{13}^u \widetilde\phi\\
    y_{21}^u \widetilde H & y_{22}^u \widetilde H & y_{23}^u \widetilde\phi\\
    0                        & 0                        & y_{33}^u \widetilde H\\
  \end{array}\right)\mathbf{u}_R
+ \overline{\mathbf{Q}}_L
  \left(\begin{array}{ccc}
    y_{11}^d H & y_{12}^d H & 0\\
    y_{21}^d H & y_{22}^d H & 0\\
    y_{31}^d \phi & y_{32}^d \phi & y_{33}^d H \\
  \end{array}\right)\mathbf{d}_R + \text{h.c.}
\end{equation}

The upper $2\times 2$ block of both quark mass matrices can be brought to diagonal form by rotations among $Q_{1,2}, u_{R,1,2}$ and $d_{R,1,2}$. Note that, after these rotations -- call them $R_{12}^{uL, dL}$, the up-type quark mass matrix has non-zero $(1,3), (2,3)$ entries, while the down-type matrix has the opposite structure: 
\begin{equation}\label{eq:mixingmatrix}
R_{12}^{ uL}.M_u.R_{12}^{ uR\dagger} =  \left(\begin{array}{ccc}
    m_u^0 & 0 & c\, m_t^0\\
    0 & m_c^0 & d\, m_t^0\\
    0 & 0 & m_t^0
\end{array}\right)
\quad  \text{and} \quad
R_{12}^{ dL}.M_d.R_{12}^{ dR\dagger} =   \left(\begin{array}{ccc}
    m_d^0 & 0 & 0\\
    0 & m_s^0 & 0\\
    a m_b^0 & b m_b^0 & m_b^0.
\end{array}\right),
\end{equation}

This structure follows automatically from the $\U$ charge of $\phi$. Given this structure, bringing the mass matrices to fully diagonal form can be accomplished by redefinitions among left handed fermions between $Q_3$ and $Q_{1,2}$ and redefinitions among the right handed down quarks between $d_{R3}$ and $d_{R,1,2}$. As the kinetic terms of the three generations are not identical, these last redefinitions generically induce FCNC in $\U$ gauge interactions. These FCNC are tightly constrained, especially in the down-quark sector. However, if we impose that $a,\,b \simeq 0$ in Eq.~\eqref{eq:mixingmatrix}, all FCNC are relegated to the up-quark sector, where constraints are weaker. In this circumstance, a viable ($3 \times 3$) CKM matrix is still generated, and one can show that the elements of the up-quark matrix in Eq.~\eqref{eq:mixingmatrix} are proportional to the CKM elements $c\,\sim V_{ub}$ and $d\,\sim V_{cb}$~\cite{1705.01822}. 

We emphasize that the choice $a \sim b \sim 0$ is not demanded by the setup, but is a phenomenological constraint. Accepting this constraint, we can work out the fermion mass basis interactions with $X$. The only place where FCNC $\U$ interactions occur is with left handed up-quarks. Specifically, expanding out the kinetic term and performing the rotations described above to go to the mass basis:
 \beq 
\label{eq:Xqq}
\Lc_{_{ XQQ}}^{\text{tree}} \simeq  \frac{\gx}{3} \bar u_L
\begin{pmatrix}
    V_{ub}^2  &  V_{ub}V_{cb}  & V_{ub} \\
    V_{ub}V_{cb} &  V_{cb}^2 & V_{cb}\\
    V_{ub}       &  V_{cb}      & 1
\end{pmatrix} 
\gamma^\mu u_LX_\mu
+ \frac{\gx} {3}\bar d_L
\begin{pmatrix}
    0     & 0  & 0 \\
    0     & 0  & 0    \\
    0 & 0  & 1
\end{pmatrix}\gamma^\mu d_L X_\mu.
\eeq
There are no off-diagonal terms present in the $u_R$, $d_R$, or leptonic interactions with $X$, so they all have the same for as the $d_L$ interaction in Eq.~\eqref{eq:Xqq}.


\bibliography{BLthird}
\bibliographystyle{JHEP}
\end{document}